\documentclass[twocolumn]{aastex62}
\usepackage{natbib}
\usepackage{gensymb}
\usepackage{amsmath}
\usepackage{enumitem}
\usepackage{graphicx}
\usepackage{multirow}
\usepackage[normalem]{ulem}
\usepackage{soul}
\usepackage{tabularx,booktabs}
\usepackage[caption=false]{subfig}
\newcommand{\Flux}{\rm MTOT}
\begin{document}

\title{\Large{Application and interpretation of deep learning for identifying pre-emergence magnetic-field patterns}}

\correspondingauthor{Dattaraj B. Dhuri}
\email{dattaraj.dhuri@tifr.res.in}

\author{Dattaraj B. Dhuri}
\affiliation{Department of Astronomy and Astrophysics, Tata Institute of Fundamental Research, Mumbai, India 400005}

\author{Shravan M. Hanasoge}
\affiliation{Department of Astronomy and Astrophysics, Tata Institute of Fundamental Research, Mumbai, India 400005}

\author{Aaron C. Birch}
\affiliation{Max-Planck-Institut f\"ur Sonnensystemforschung, G\"ottingen, Germany}

\author{Hannah Schunker}
\affiliation{Max-Planck-Institut f\"ur Sonnensystemforschung, G\"ottingen, Germany}
\affiliation{School of Mathematical and Physical Sciences, The University of Newcastle, New South Wales, Australia}

\begin{abstract}
Magnetic flux generated within the solar interior emerges to the surface, forming active regions (ARs) and sunspots. Flux emergence may trigger explosive events - such as flares and coronal mass ejections and therefore understanding emergence is useful for space-weather forecasting. Evidence of any pre-emergence signatures will also shed light on sub-surface processes responsible for emergence. In this paper, we present a first analysis of emerging ARs from the \emph{Solar Dynamics Observatory/Helioseismic Emerging Active Regions} (SDO/HEAR) dataset \citep{Schunker2016} using deep convolutional neural networks (CNN) to characterize pre-emergence surface magnetic-field properties. The trained CNN classifies between pre-emergence (PE) line-of-sight magnetograms and a control set of non-emergence (NE) magnetograms with a {\it True Skill Statistic} (TSS) score of ~ $\sim 85\%$, 3~h prior to emergence and $\sim 40\%$, 24~h  prior to emergence. Our results are better than a baseline classification TSS obtained using discriminant analysis of only the unsigned magnetic flux. We develop a network pruning algorithm to interpret the trained CNN and show that the CNN incorporates filters that respond positively as well as negatively to the unsigned magnetic flux of the magnetograms. Using synthetic magnetograms, we demonstrate that the CNN output is sensitive to the length-scale of the magnetic regions with small-scale and intense fields producing maximum CNN output and possibly a characteristic pre-emergence pattern. Given increasing popularity of deep learning, techniques developed here for interpretation of the trained CNN --- using network pruning and synthetic data --- are relevant for future applications in solar and astrophysical data analysis.
\end{abstract}
\section{Introduction \label{sec:Intro}} 
Magnetic flux in the Sun, generated by the dynamo operating within the interior, rises to the visible surface, the photosphere, and  further into the solar atmosphere \citep{Cheung2014,Stein2012}. The rising flux appears at the photosphere in the form of large-scale ($\sim 10-100~\textrm{Mm}$) structures of concentrated magnetic flux ($\sim \textrm{kG}$), known as sunspots and active regions (ARs). In the low-$\beta$ solar atmosphere and corona, these flux tubes expand significantly and form gigantic loop structures rooted in the ARs. The rising magnetic flux also pumps into ARs non-potential energy that powers explosive events such as flares and coronal mass ejections (CMEs) that can lead to severe space weather consequences \citep{Shibata2011,Eastwood2017}. Knowledge of the flux emergence mechanism in ARs is thus useful for gaining warning time for space-weather forecasting. Additionally, a comprehensive picture of the solar dynamo includes processes responsible for generation as well as transport of the magnetic flux \citep{Cheung2014}. Understanding the formation and evolution of ARs is therefore necessary for constraining solar dynamo models \citep{Schunker2016,Birch_2012,Cameron2016}.

The physical mechanism that leads to the appearance of new ARs on the photosphere, henceforth referred to as ``emergence", are not fully known \citep{Leka_2012,Schunker2016,Cheung2014}. Broadly, two categories of emergence scenarios are considered in the literature. In the first scenario, flux tubes may form deep in the convection zone and rise intact to emerge on the surface \citep{Fan2009}. Alternatively, small-scale flux tubes, which may be generated within the bulk of the convection zone or in the near-surface layers, coagulate and form ARs on the surface \citep{Brandenburg_2005,Brandenburg2014}.
Case-studies for detecting sub-surface pre-emergence signatures, e.g. using helioseismology \citep{Birch_2010}, may not be successful because of the signal being weak in comparison to the uncertainties in detection. Therefore, for unambiguous detection of pre-emergence signatures, a statistical study of a large sample of emerging ARs is in order. A statistical study, such as \cite{Komm2009,Komm2011}, is expected to capture common characteristics of emerging ARs which may be universal to the emergence mechanism \citep{Schunker2016}.

\citet{Leka_2012,Birch_2012,Barnes_2014} carried out an extensive survey of emerging ARs, henceforth referred to as the LBB survey, using helioseismic holography. Analyzing SOHO/MDI line-of-sight magnetograms and Global Oscillation Network Group (GONG) dopplergrams, the LBB survey detected weak converging flows and reduced travel-times to distinguish emerging active regions (EARs) from a set of control regions (CRs) which did not show emergence. The study also identified pre-emergence magnetic field signal correlated with converging flows in the EAR population. Although a causal relationship between the converging flow and the pre-emergence magnetic field signal could not be established, the average unsigned radial magnetic field of EARs was demonstrated to be the leading discriminator between the EAR and CR populations, yielding True Skill Statistics (Peirce Skill Score) \citep{PEIRCE453,bobraflareprediction} of $\sim 50\%$, 3~h prior to the emergence. This warrants further investigation of pre-emergence surface magnetic-field characteristics.

Motivated by the results of the LBB survey, here we perform comprehensive statistical analysis to attempt to uncover spatio-temporal patterns associated with pre-emergence surface magnetic fields. Surface magnetic-field data of superior spatial and temporal resolution is available from {\it Helioseismic and Magnetic Imager} (HMI) \citep{Schou2012} on-board {\it Solar dynamics Observatory} (SDO) \citep{Pesnell-etall2012}. Following the LBB survey, \citet{Schunker2016} assembled the {\it Helioseismic Emerging Active Region} (SDO/HEAR) dataset, also comprising pre-emergence magnetograms of emerging ARs (EARs) and control regions (CRs). Although originally designed for measuring sub-surface pre-emergence activity, the SDO/HEAR dataset is also useful to study surface magnetic-field properties of emerging ARs \citep{Schunker2016,Schunker2019}. 
The \emph{SDO/HEAR} dataset comprises some $\sim 200$ samples of EARs and CRs each, with $60\degree \times 60\degree$ spatial extent and tracked over a duration of up to one week before emergence. Detecting spatio-temporal correlations and patterns over such a large dataset of magnetograms is statistically challenging. With the advent of machine learning (ML), however, advanced algorithms to analyze and classify images to detect complex correlations are available \citep{hastie01statisticallearning}. In particular, deep convolutional neural networks (CNNs) have proven immensely successful for visual recognition tasks, such as image and video classification, by accurately characterising structural information from the images \citep{Goodfellow2016,Krizhevsky2012,LeCun2015}. Here, we use CNNs to discriminate between magnetograms from EAR and CR populations and in this process, uncover associated pre-emergence surface magnetic-field characteristics.

The classification problem considered here can be naturally formulated as a supervised learning \citep{hastie01statisticallearning} problem, where the CNN is trained using the dataset of magnetograms labeled as pre-emergence (PE) and non-emergence (NE). The CNN is trained to optimize the {\it True Skill Statistics} (TSS) score for classification of PE and NE magnetograms taken at different pre-emergence times. CNNs, and deep neural networks in general, may be trained for challenging tasks. However, it is difficult to interrogate the operation of neural networks into comprehensible components and thus interpret their performance. For the present work, the interpretation of the trained network is important for obtaining quantifiable information about pre-emergence surface-magnetic-field characteristics. Our analysis in this work therefore focuses on understanding the performance of the network using synthetic magnetograms and advanced techniques such as functionality based network pruning \citep{Cun90,Han2015,Qin2018,Frankle2019}. 

The paper is organized as follows. In Section~\ref{sec:Data}, we describe the processing of EARs and CRs in the \emph{SDO/HEAR} dataset for deep learning analysis using a CNN. In Section~\ref{sec:Methods}, we explain the CNN architecture  and training methodology. In Section~\ref{sec:Results}, we describe in detail the results of the classification of pre-emergence (PE) and non-emergence (NE) magnetograms and compare with earlier work using discriminant analysis. In particular, we explain the working of the CNN using the network-pruning algorithm developed here to facilitate the CNN interpretation. We detail various statistical analyses performed to comprehend the CNN performance. We also explain probing of the trained CNN using synthetic magnetograms. In Section~\ref{sec:Discussion}, we summarize our findings.
\begin{figure}[b]
\centering
\includegraphics[width=0.45\textwidth,trim={2cm 1cm 2cm 2cm},clip]{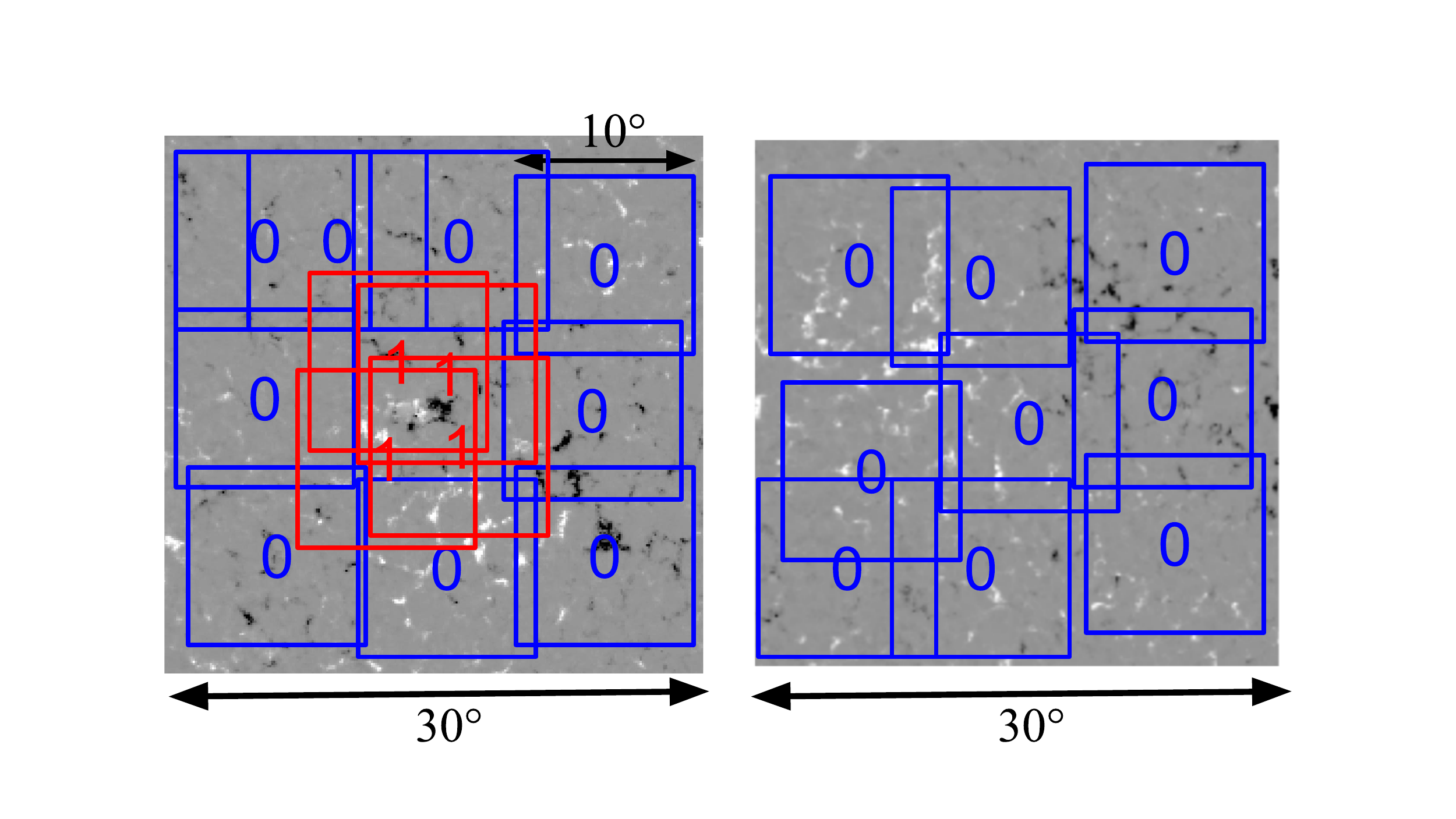}
\caption{Sub-sampling pre-emergence (PE, {\it red}) and non-emergence (NE, {\it blue}) magnetogram segments from the EARs ({\it left}) and CRs ({\it right}) in the \emph{SDO/HEAR} dataset \citep{Schunker2016}. The $10\degree \times 10\degree$ segments are randomly sub-sampled from the original $30\degree \times 30\degree$ EARs and CRs. Sub-sample locations are randomly selected for each EAR/CR, thus they are different for different EARs and CRs. We constrain the sub-sampling to ensure that the emergence region (the central $5\degree \times 5\degree$ of EARs) is fully included (fully excluded) within PE (NE) magnetogram segments. We sub-sample NE magnetogram segments from CRs as well as non-central region of EARs.  PE and NE segments are Postel projections of size $85 \times 85$ pixels, with a plate scale of 1.4~Mm per pixel, i.e. same as the original EARs/CRs. We label PE and NE segments as 1 and 0 respectively for supervised learning using the deep neural network.}
\label{fig:subsampling}
\end{figure}

\section{Data \label{sec:Data}} 
We use the \emph{SDO/HEAR} dataset comprising pre-emergence line-of-sight magnetograms of active regions observed by SDO/HMI between May 2010 and July 2014 \citep{Schunker2016}. EARs in the dataset are selected to be emergences into relatively quiet Sun to minimize difficulty in distinguishing signatures of emerging flux from any pre-existing magnetic field. Each EAR is paired with a control region (CR) which does not show emergence. Each CR is co-spatial with the corresponding EAR, in terms of the latitude and distance from the central meridian, and is separated by no more than two solar rotation periods. The \emph{SDO/HEAR} dataset, although originally prepared for a helioseismology survey, is also useful for studying pre-emergence surface magnetic-field properties \citep{Schunker2016}. The dataset provides Postel-projected EAR and CR maps covering $60\degree \times 60\degree$ about the emergence location. However, following the LBB survey, we restrict our analysis to the central $30\degree \times 30\degree$ region. This region is projected onto a $256 \times 256$-pixel grid with a plate scale of 1.4~Mm per pixel. In the following, we explain in detail the problem at hand to classify between $10\degree \times 10\degree$  pre-emergence (PE) and non-emergence (NE) magnetogram segments sub-sampled from the original $30\degree \times 30\degree$ magnetograms of EARs and CRs in the \emph{SDO/HEAR} dataset (see Figure~\ref{fig:subsampling}). We use the following definitions of the magnetograms referred to in this work (see Table~\ref{tab:Dataset}). 

\begin{itemize}[nolistsep,noitemsep]
    \item EARs - Emerging active regions of size $30\degree\times 30\degree$ taken from the \emph{SDO/HEAR} dataset.
    \item CRs - Control regions of spatial extent $30\degree\times 30\degree$ not showing emergence, taken from the \emph{SDO/HEAR} dataset.
    \item PEs - Pre-emergence magnetograms of spatial extent $10\degree\times 10\degree$ randomly sub-sampled from the central regions of EARs, as shown in Figure~\ref{fig:subsampling} (red).
    \item NEs - Non-emergence magnetograms of spatial extent $10\degree\times 10\degree$ randomly sub-sampled from CRs and non-central regions of EARs, as shown in Figure~\ref{fig:subsampling} (blue).
\end{itemize}
\begin{table}[b] 
\centering
\begin{tabular}{c|cc|cc}
\toprule
time before & EARs & CRs & PEs & NEs\\
 emergence &\multicolumn{2}{c|}{$30 \degree \times 30 \degree$} &\multicolumn{2}{c}{$10 \degree \times 10 \degree$} \\
(h)  & \multicolumn{2}{c|}{(original)}  & \multicolumn{2}{c}{(sub-sampled)} \\
 \midrule
-3.2 & 178 & 180 & 17800 & 18080 \\
-8.5 & 174 & 165 & 17400 & 17115  \\
 -13.9 & 173 & 169 & 17300 & 17269 \\
 -19.2 & 172 & 167 & 17200 &  17117 \\
-24.5 & 158 & 158 & 15800 & 15958 \\
 \bottomrule
\end{tabular}
\caption{The {\it SDO/HMI survey of emerging active regions} (SDO/HEAR) dataset: We consider emerging active regions (EARs) and control regions (CRs) between May 2010 and July 2014. We sub-sample pre-emergence (PE) and non-emergence (NE) segments from original EARs and CRs to furnish sufficient data for training a deep neural network.}
\label{tab:Dataset}
\end{table}
\begin{figure*}[t] 
\centering
\includegraphics[width=\textwidth,trim={4cm 0.2cm 4cmemergence 0.5cm},clip]{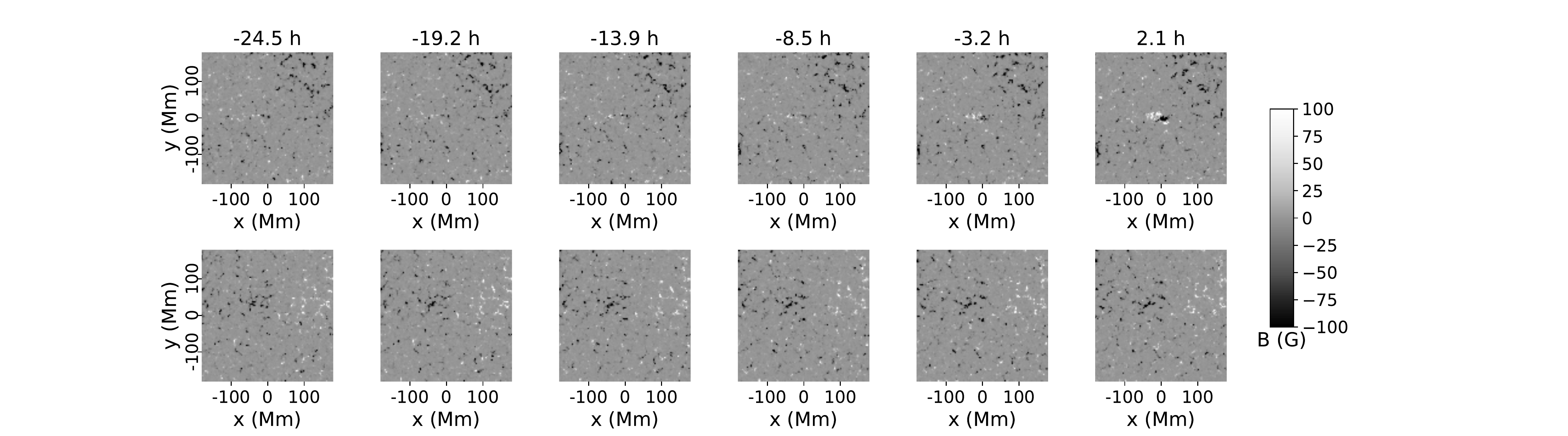}
\caption{Example emerging active region (EAR, {\it top}) and control region (CR, {\it bottom}) from the \emph{SDO/HEAR} dataset \citep{Schunker2016}. Temporally averaged pre-emergence Postel-projected magnetograms for EAR 11776 and corresponding CR are shown. The magnetograms are spread over $30 \degree \times 30 \degree$ about the emergence location and projected onto a $256 \times 256$-pixel grid with a plate scale of $1.4~\textrm{Mm}$ per pixel. The emergence location is confined to the central $10 \degree \times 10 \degree$ region. The bipolar signature of the EAR makes a clear appearance as the emergence time approaches. Pre-emergence surface magnetic fields of EARs and CRs are otherwise difficult to distinguish. The color scale is saturated at $\pm~100~{\rm G}$.}
\label{fig:EAR-CR_Example}
\end{figure*}
\begin{figure*}[t]
\centering
\includegraphics[width=\textwidth,trim={4cm 0.2cm 4cm 0.5cm},clip]{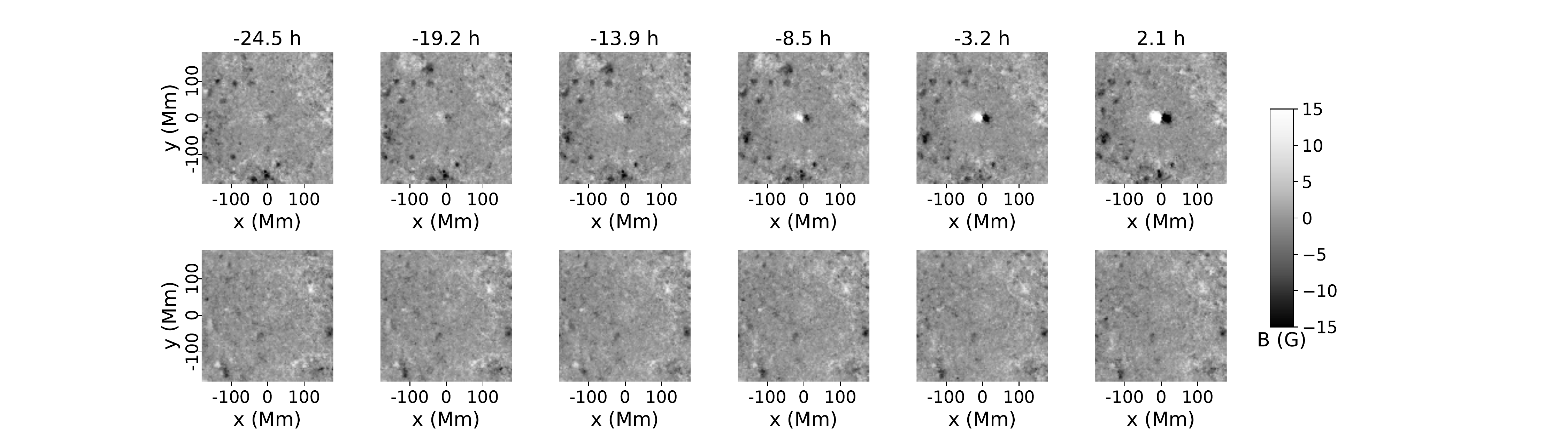}
\caption{Average emerging active region (EAR, {\it top}) and control region (CR, {\it bottom}) from the \emph{SDO/HEAR} dataset \citep{Schunker2016}. The magnetograms from the dataset at each pre-emergence time are averaged after accounting for Hale's and Joy's laws. Patterns in the pre-emergence bipolar surface magnetic field may be seen distinctly in EARs. The deep convolutional neural network (CNN) may learn this faint bipolar field signature and correlated patterns. The color scale is saturated at $\pm~15~{\rm G}$.}
\label{fig:EAR-CR_Averaged}
\end{figure*}

The \emph{SDO/HEAR} dataset is compiled following several criteria to facilitate characterization of EARs and CRs using statistical analysis \citep{Schunker2016}. Only those EARs are selected which (1) are visible in the continuum, (2) reach total area of at least $\rm{10~\mu H\ or\ 30~Mm^2}$, and (3) are within $\pm 50\degree $ of the central meridian at the time of emergence $t_s^{\rm NOAA}$, as defined by National Oceanic and Atmospheric Administration (NOAA). Consistent with the LBB survey, the emergence time $t_0$ of an EAR is defined as the time when the absolute flux of a EAR is $10\%$ of the maximum observed value within a $\rm{36~h}$ interval following $t_s^{\rm NOAA}$. Moreover, EARs are selected only when the absolute flux rises monotonically after the emergence from a low, steady value.  Following these restrictions, $182$ EAR samples between May 2010 and July 2014 are included in the \emph{SDO/HEAR} dataset. CRs in the dataset are selected to match the spatial distribution of EARs. This is achieved by selecting a CR co-spatial with each EAR and also from the identical phase of the solar cycle. The absolute difference between the average pre-emergence magnetic field within the central $10\degree \times 10\degree$ of a CR and associated EAR is restricted to be less than 10~G or $1.4\times10^{21}~{\rm Mx}$ (see Figure 3 in \cite{Schunker2016}). Importantly, the CRs do not show marked increase in the absolute flux similar to emergence. CRs are assigned a mock-emergence time when their Stonyhurst coordinates \citep{Thompson2005} are identical to the corresponding EARs, at the time of emergence $t_0$.  

The EARs and CRs are tracked up to a consecutive two-week duration, up to one week on either side of emergence, at the Carrington rotation rate. The SDO/HMI line-of-sight magnetograms are available at a cadence of 45~s. The magnetograms are temporally averaged over intervals of 410.25~min (547 frames) and spaced by 320.25~min (427 frames) to smooth out temporal fluctuations. Thus, there is an overlap of 90~min (120 frames) between consecutive temporal-averaging intervals \citep[see Figure 5 in][]{Schunker2016}.  Consistent with the LBB survey, we only consider pre-emergence temporally averaged magnetograms up to a day prior to emergence. The number of EAR and CR samples considered in our analyses, at different pre-emergence times, are listed in Table~\ref{tab:Dataset}. We include only those EARs and CRs for which magnetogram samples are available at all pre-emergence times and therefore the number of EARs and CRs is not exactly equal.  

Figure~\ref{fig:EAR-CR_Example} shows time-averaged pre-emergence magnetograms of an example EAR and corresponding CR. As the emergence time approaches, we see the early appearance of a bipolar field within the central $10\degree \times 10\degree$ of EARs. Apart from that, we see no obvious difference between EARs and CRs prior to emergence. The signature of the pre-emergence bipolar field becomes evident when we consider the average of all EARs and CRs in the dataset, shown in Figure~\ref{fig:EAR-CR_Averaged}, after accounting for Hale's polarity law and Joys's law \citep{Schunker2016}. To account for Hale's law, we reverse the sign of the magnetic field for the magnetograms in the southern hemisphere. To account for Joy's law, we flip the magnetograms in the southern hemisphere in the latitudinal direction, i.e $+y$ direction points away from the equator. For the average EAR, the pre-emergence field within the central $10\degree \times 10\degree$ region may be distinctly seen, even up to a day prior to the emergence. As mentioned earlier, this pre-emergence magnetic field was a leading factor in the discriminant analysis performed in the LBB survey to differentiate the EAR and CR samples. The present analysis, using deep learning, is expected to identify this pre-emergence field as well as other correlated patterns.

As in the case of the LBB survey, the \emph{SDO/HEAR} dataset used in this analysis is specifically designed for examination of pre-emergence signatures using helioseismology and surface magnetic-field properties. The EARs and CRs are selected with a bias of prior knowledge of where emergence occurred. Therefore, our work is not an attempt of ``forecasting" emergence. Rather, the focus is on characterizing EARs and CRs based on their surface magnetic-field properties and understanding the process of emergence. The \emph{SDO/HEAR} dataset is well suited for studying spatio-temporal characteristics of the pre-emergence surface magnetic field, as demonstrated in \citet{Schunker2016}.

A major obstacle for our analysis is that the number of EAR samples is indeed small in the \emph{SDO/HEAR} dataset for training a deep neural network. It is difficult to know a priori the exact number of samples required for successfully training a deep neural network. The number of parameters in such a network are $\mathcal{O}(100,000)$ - many of these may be redundant, but for successful training, thousands of examples from each category (class) may be necessary \citep{Goodfellow2016}. In order to have sufficient number of examples for training, we consider pre-emergence magnetograms of smaller size in our analysis. These are sub-sampled from the original $30\degree \times 30\degree$ Postel-projected maps. By design, the emergence is restricted to the central $10\degree \times 10\degree$ area of EARs in the \emph{SDO/HEAR} dataset. Thus, we obtain magnetogram segments of $10\degree \times 10\degree$ selected randomly from the original EAR and CR magnetograms. Because of the random selection the sub-sampled segments are different geometric regions in different EARs and CRs. We constrain the random selection to ensure that the sub-sampled $10\degree \times 10\degree$ segments either fully contain or fully exclude the emergence location, i.e., segments that contain the emergence location even partially are disallowed (see Figure~\ref{fig:subsampling}). However, there is no minimum distance between two random PE or NE from an EAR or CR i.e. there may be significant overlap. Note that the NE segments are sub-sampled from the CRs as well as from the non-central region of the EARs. Also, the NE segments in the training set could include regions where flux does emerge, that were not classified as PE regions in the \emph{SDO/HEAR} survey. This could increase the chances of the machine learning algorithm returning false negatives. The PE and NE segments are now Postel-projected maps on $85 \times 85$ pixel grids. The plate scale of the magnetogram segments is 1.4~Mm per pixel, i.e. similar to EARs and CRs. We subsample exactly 100 PEs from the central part of each EAR and approximately 50 NEs each from the non-central part of each EAR and from each CR. Table~\ref{tab:Dataset} lists the total number of PE and NE segments used for analysis.   We train a deep neural network, a machine learning algorithm, to classify between these PE and NE segments labeled as class 1 and class 0 respectively.
\begin{figure}[b]
\centering
\includegraphics[width=0.45\textwidth,trim={3cm 6cm 3cm 3cm},clip]{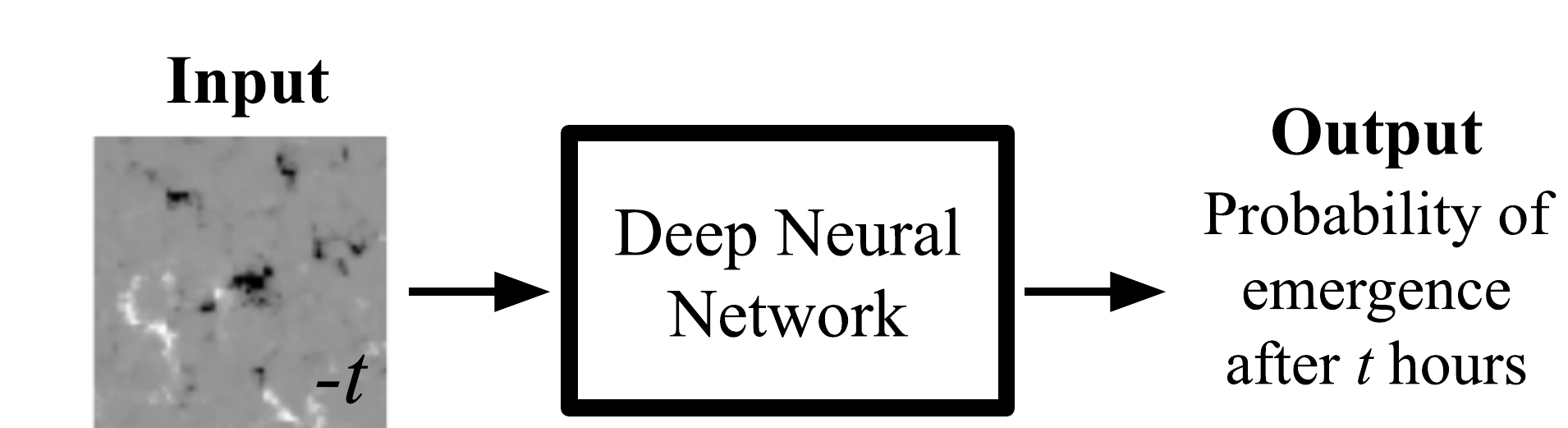}
\caption{Schematic for deep learning classification of pre-emergence (PE) and  non-emergence (NE) magnetogram segments.}
\label{fig:Schematic}
\end{figure}

\section{Methods} \label{sec:Methods}
Machine learning is a set of algorithms applied to analyze patterns and correlations in large, high-dimensional datasets without being explicitly programmed for it. Here, we apply supervised learning, where the machine is trained using a known set of input-output pairs from a part of the available data \citep{hastie01statisticallearning}. PE and NE magnetograms, at a given pre-emergence time $t$, are inputs and class labels 1 and 0 respectively are the desired outputs (see Figure~\ref{fig:Schematic}). The machine's parameters, i.e. weights and biases (see Appendix~\ref{app:NNs}), are tuned during the training process when it learns the probability distribution of the inputs. The trained machine is used to make predictions on the remaining part of the data which is called validation or test data. We use deep convolutional neural networks (CNN) as our method of choice for the classification of PE and NE magnetograms \citep{Krizhevsky2012,LeCun2015,Goodfellow2016}. CNNs have been used extensively for analysis of images and proven widely successful. 

\begin{figure}[t]
\centering
\includegraphics[width=0.45\textwidth,trim={4cm 3cm 4cm 3cm},clip]{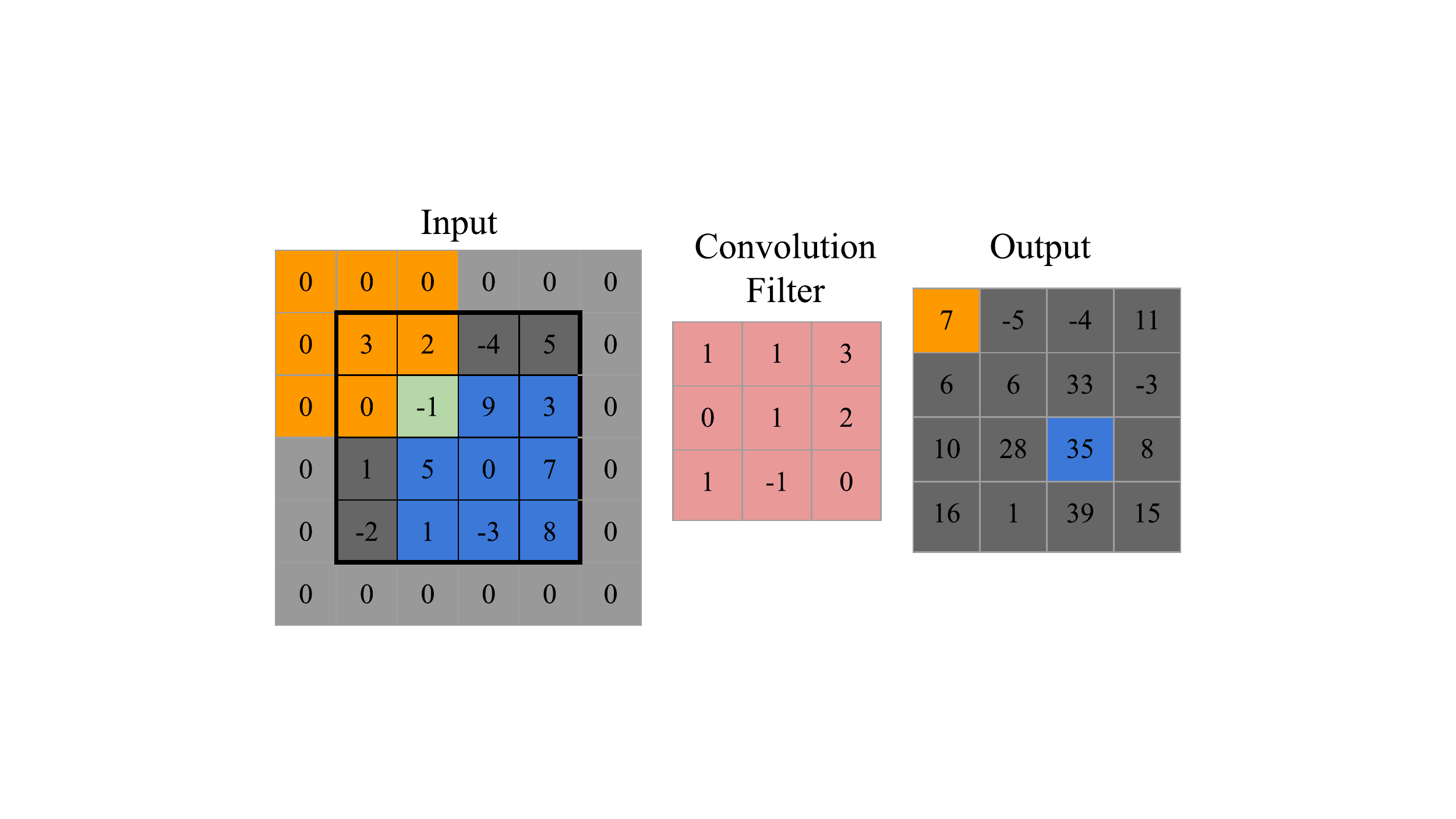}
\caption{Operation of a $3 \times 3$ convolutional filter with the identity activation function. For illustration purposes, a $4 \times 4$ input is considered. Zero padding is applied to the input to extract the output of the convolution operation of the same size as that of the input. The $3 \times 3$ convolution filter consists of nine neurons, i.e. nine weights (see Appendix~\ref{app:NNs}), as shown. The filter slides over each $3 \times 3$-pixel sub-region from the input in order. Assuming that the bias $b$ of the filter is $0$, the identity activation of the neurons yields the output $\sum w_ix_i$ for each sub-region. The outputs for the sub-regions shaded in orange and blue color are respectively highlighted.}
\label{fig:ConvFilter}
\end{figure}

\begin{figure*}[t]
\centering
\includegraphics[width=0.8\textwidth,trim={0cm 5cm 0cm 2cm},clip]{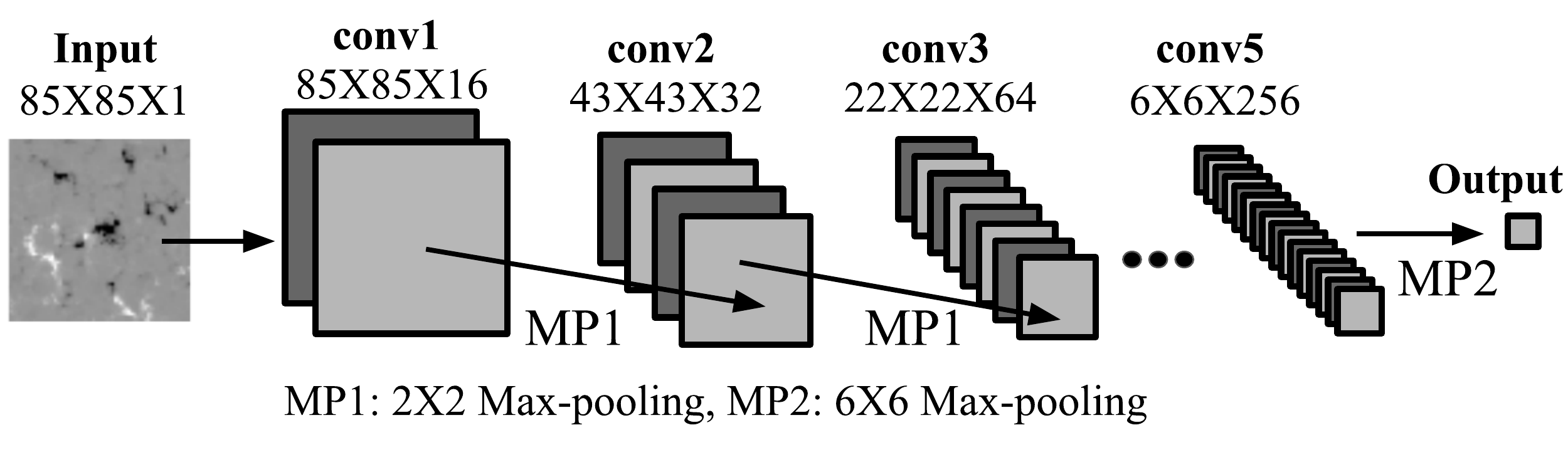}
\caption{Fully convolutional neural network (CNN) \citep{LeCun2015,Goodfellow2016} used for classification of pre-emergence (PE) and non-emergence (NE) magnetogram segments. The network architecture is based on VGGNet \citep{Simonyan2014} and incorporates $3\times3$ convolutional filters (Figure~\ref{fig:ConvFilter}) and $2\times2$ max-pooling filters. The max-pooling filters downsize the output of each convolutional layer by a factor of two. Similar to the VGGNet, the number of convolutional filters double in successive convolutional layers. Unlike the VGGNet, there are no fully connected layers of neurons connecting the final convolutional layer to the output. Instead, the  final convolution layer is connected to the output via $6 \times 6$ max-pooling operation. This facilitates interpretation of the trained neural network. }
\label{fig:CNN}
\end{figure*}

The CNN consists of layers of neurons with 2D convolutional filters, particularly designed for analyzing images.  A convolutional filter of $K \times K$ neurons slides over input magnetograms (Figure~\ref{fig:ConvFilter}), transforming a $K \times K$-pixel sub-region at a time according to the neuron activation function (Eq.~\ref{eq:neuronAct}). Unlike typical image classification problems, the magnetograms contain inputs with both positive and negative values representing opposite magnetic polarities. Therefore, we use the identity activation  function  for  CNN filters,  i.e.,  every neuron in the CNN filter outputs $y=\sum_i{w_i x_i + b}$,  where $\textbf{x}$ are  inputs  to  the  neuron with  weights $\textbf{w}$ and $b$ is the bias of the filter (Appendix~\ref{app:NNs}). The identity activation  function  ensures  that positive and negative (magnetic-field) values from the input are treated equally by the network. In addition to the convolutional layers, max-pooling layers are also used in CNNs.  A max-pooling layer of size $M \times M$ extracts the maximum value  from an $M \times M$-pixel sub-region,  thus  downsampling  the input by a factor of $M$.  A combination of convolutional and pooling layers, placed at increasingly deeper levels, are sensitive to features of increasing length scales of the original image. We use a CNN architecture based on VGGNet \citep{Simonyan2014} which uses $3 \times 3$ convolutional filters  and $2 \times 2$ max-pooling filters (see Figure~\ref{fig:CNN}). The VGGNet architecture includes fully connected (FC) layers to connect the final convolutional layer to the output. To facilitate interpretation of the trained CNN by analysis of the features learned by convolutional filters in the final layer, we incorporate a CNN without any FC layers i.e. a fully convolutional neural network \citep{ShelhamerFCN2017}. The 5-layer deep CNN that we use reduces the original $85 \times 85$-pixel input to $6 \times 6$ pixels via max-pooling operations. 

During training, error in the predicted output, as measured by a loss function, is minimized by tuning the weights and biases of the network via stochastic gradient descent (see Appendix~\ref{app:NNs}). Here, we use binary cross-entropy loss function $L_{\rm CE}$ (Eq.~\ref{eq:LossFunction}). Training a neural network also involves fine tuning the network hyper-parameters. The hyper-parameters associated with stochastic gradient descent for training include learning rate $lr$ and batch size $N_{\rm BS}$. The learning rate determines the step of the gradient descent. Too small a learning rate slows down the training and too large a learning rate diverges the loss function value $L_{\rm CE}$ (Eq.~\ref{eq:LossFunction}) \citep{hastie01statisticallearning}. Batch size determines the number of training examples considered for the gradient descent at each iteration. We tune learning rate, batch size and depth of the CNN (number of convolutional layers) to optimize the classification of the PE and NE magnetograms as measured by True Skill Statistics (TSS, see below).

Subsequent to the final convolutional layer, we perform a $6 \times 6$ max-pooling operation to reduce the output to the one neuron in the final layer. We choose the activation of the output neuron as a sigmoid function that yields a value between 0 and 1 \citep{hastie01statisticallearning}. This corresponds to the probability of the input magnetogram containing the emergence. The predicted classes are labeled as positive (1) and negative (0), by thresholding the CNN output at 0.5. The CNN output thus falls in one of the following categories. 
\begin{itemize}[noitemsep]
\item True Positives (TPs) - Sub-population of emerging magnetic field segments (PE) accurately classified as positive (1).
\item True Negatives (TNs) - Sub-population of non-emerging magnetic field segments (NE) accurately classified as negative (0).
\item False Positives (FPs) - Sub-population of non-emerging magnetic field segments (NE) inaccurately classified as positive (1).
\item False Negatives (FNs) - Sub-population of emerging magnetic field segments (PE) inaccurately classified as negative (0).
\end{itemize}
Note that the FPs categorized here as such may still include true emergences that do not fall under the \emph{SDO/HEAR} criteria.

Following the LBB survey, we use the Peirce Skill Score or True Skill Statistics (TSS) to measure the machine performance \citep{Barnes_2014,PEIRCE453,bobraflareprediction}. The TSS is defined as ${\rm TSS} = TP/(TP+FN) - FP/(FP+TN)$. The TSS is 0 for both random and unskilled predictions and 1 for the perfect classification.

\section{Results \label{sec:Results}} 
\subsection{Training}
\begin{figure}[b]
\centering
\includegraphics[width=0.45\textwidth,trim={5cm 2cm 5cm 2cm},clip]{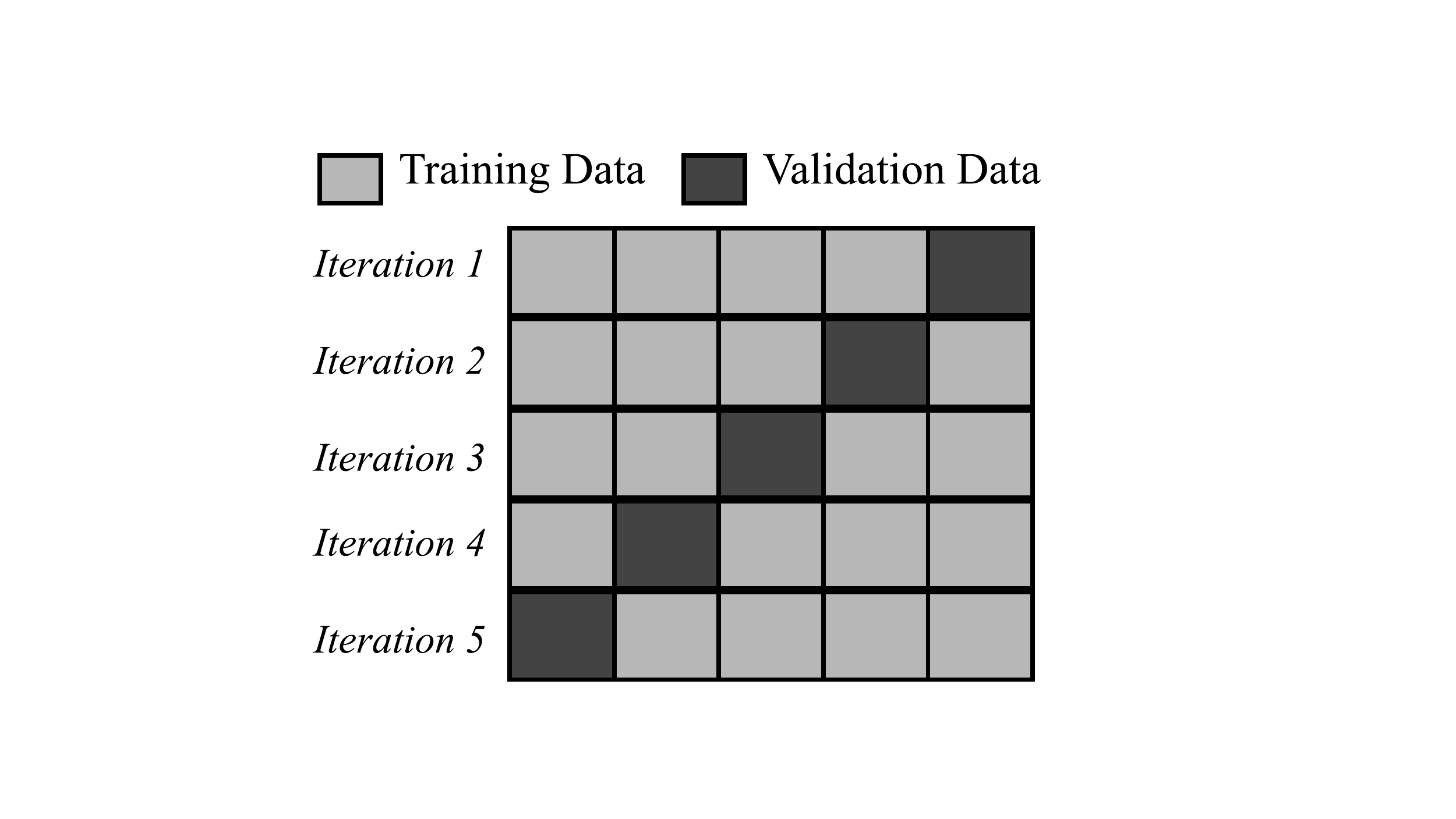}
\caption{5-fold cross-validation of the convolutional neural network for classification of pre-emergence (PE) and non-emergence (NE) magnetogram segments. The emerging active regions (EARs) and control regions (CRs) from the \emph{SDO/HEAR} dataset are randomly split in five approximately equal parts. PE and NE segments from EARs and CRs from the four parts are used for training and the remaining part is used for validation or test. This is performed five times for the five different validation sets. CNN hyper-parameters are tuned to achieve maximum average 5-fold cross-validation {\it True Skill Statistics} (TSS).}
\label{fig:CV}
\end{figure}

We train the CNN (Figure~\ref{fig:CNN}) using supervised learning \citep{hastie01statisticallearning} with PE and NE magnetograms as inputs and labels 1 and 0 respectively as outputs. Typically, the available data are split into three categories --- training, validation and testing. The training data are used to determine weights and biases of the network and validation data are used to tune the network hyper-parameters (Section~\ref{sec:Methods}). The trained machine is then used to make predictions on the test data. A well-trained network is able to reproduce the high accuracy obtained during training on test data.

Since the number of EARs and CRs available are limited (see Table~\ref{tab:Dataset}), we use 5-fold cross-validation for tuning network hyper-parameters instead of a dedicated testing set (Figure~\ref{fig:CV}) \citep{hastie01statisticallearning}. The EARs and CRs are each randomly split in five (approximately) equal parts. We use PE and NE magnetogram segments from the four parts for training and the remaining part for validation. Note that the sub-sampled PE and NE magnetogram segments from an EAR or CR are included in either the training data or the validation data and not both. We perform the training five times, with EARs and CRs from five different parts used for validation. After each training, we obtain TSS using predictions on PE and NE magnetograms from validation data. Because PE and NE magnetograms are sub-sampled randomly from EARs/CRs, PE and NE segments obtained from an EAR or CR contain overlapping regions. Therefore, while obtaining TSS only non-overlapping segments are considered. There are nine such non-overlapping $10\degree \times 10\degree$ segments from each EAR and CR (which are $30\degree \times 30\degree$) i.e. 17 NE segments for every PE segments. Thus the classification problem considered here is class-imbalanced. The network hyper-parameters are tuned to obtain the maximum average TSS value over 5-fold cross-validation. TSS is a good metric for such class-imbalanced problems.
\begin{figure}[t]
\centering
\includegraphics[width=0.40\textwidth,trim={0.2cm 0.3cm 0.2cm 0.2cm},clip]{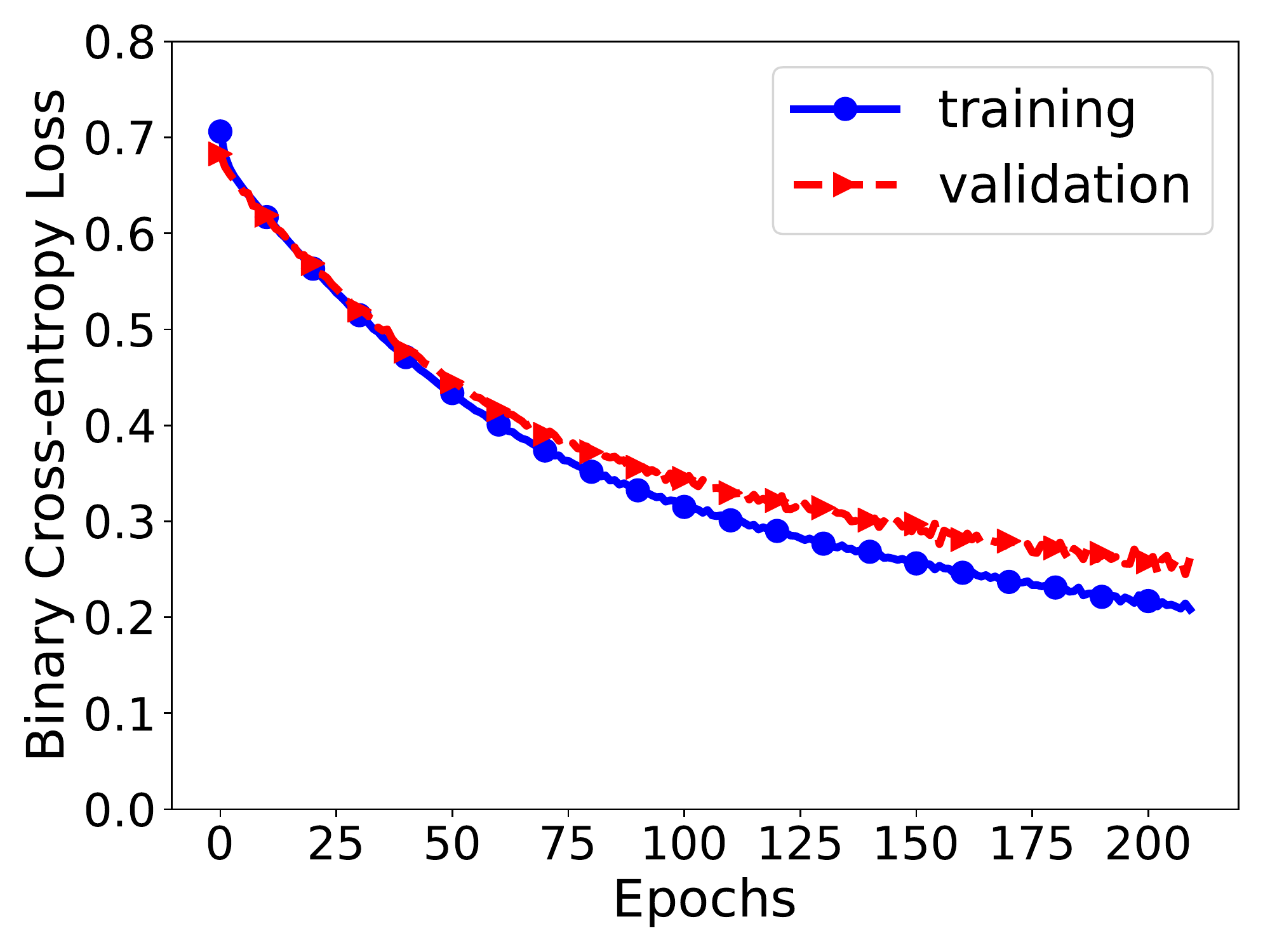}
\caption{The variation of binary cross-entropy loss $L_{\rm CE}$ (Eq.~\ref{eq:LossFunction}) or misfit between true and predicted outputs for training and validation data as a function of training iterations or epochs. The training and validation loss is approximately equal and both monotonically decrease as the training progresses indicating that parameters of the CNN are not overfitted.}
\label{fig:trainingLoss}
\end{figure}
\begin{figure*}[t]
\centering
\subfloat
{
\includegraphics[width=0.45\textwidth,trim={0cm 0cm 0cm 0cm},clip]{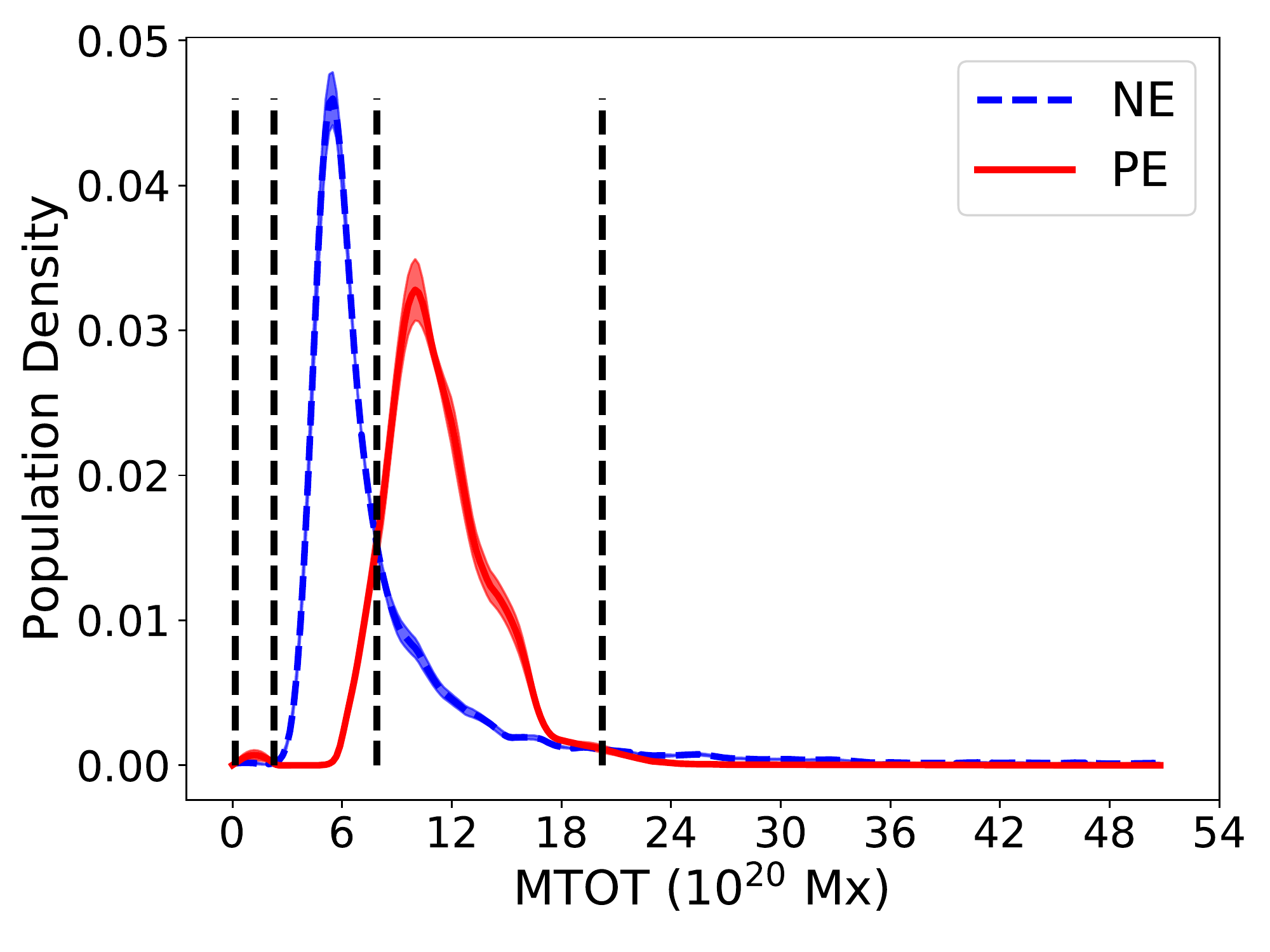}
\label{fig:DiscriminantAnalysis}
}
\subfloat{
\includegraphics[width=0.45\textwidth,trim={0cm 0cm 0cm 0cm},clip]{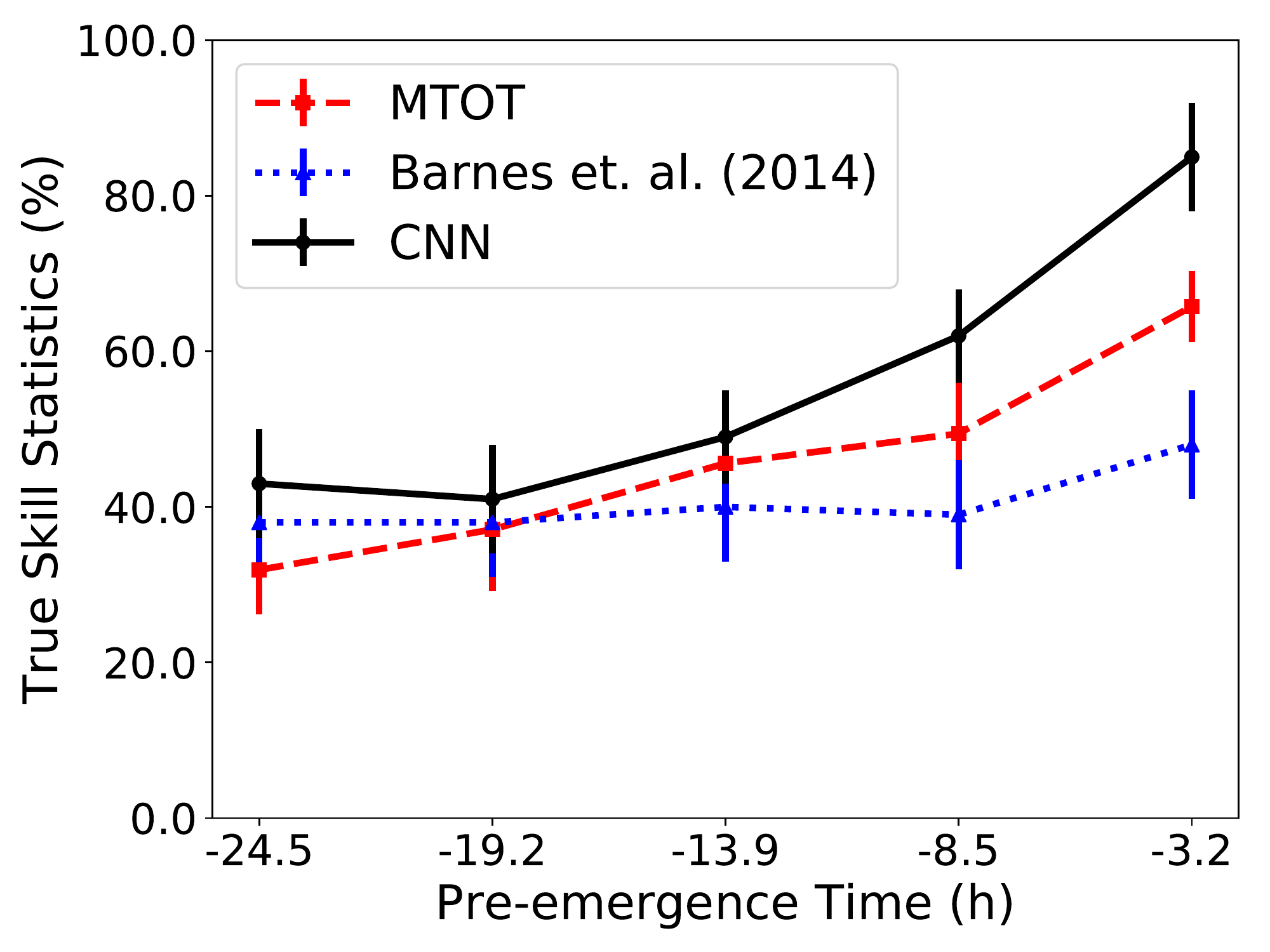}
\label{fig:TSSComparison}
}
\caption{ {\it Left:} Discriminant analysis of pre-emergence (PE) and non-emergence (NE) magnetogram segments, taken 3.2~h before emergence, based on the total unsigned line-of-sight magnetic flux $\Flux=\sum \vert B_{\rm LOS} \vert~dA$. The {\it True Skill Statistics} (TSS) achieved is $\sim 65\%$, significantly lower than the TSS obtained using the CNN. Dashed lines show discriminant boundaries, where PE and NE population densities are equal. {\it Right:} Comparison of TSS achieved for classification of PE and NE magnetogram segments using the CNN (this work) and discriminant analysis using $\Flux$ of PE and NE magnetograms taken at different pre-emergence times. Also shown are results reported in the LBB survey \citep{Barnes_2014} for classification of EARs and CRs using the discriminant analysis applied to average unsigned radial magnetic field. The CNN outperforms the discriminant-analysis classification, most significantly for magnetograms at pre-emergence times -3.2~h and -8.5~h. The $1\sigma$ error bars are shown.}
\label{fig:DA}
\end{figure*}
\begin{figure*}[t]
\centering
\subfloat
{
\includegraphics[width=0.50\textwidth,trim={2.3cm 0.8cm 2.7cm 0.8cm},clip]{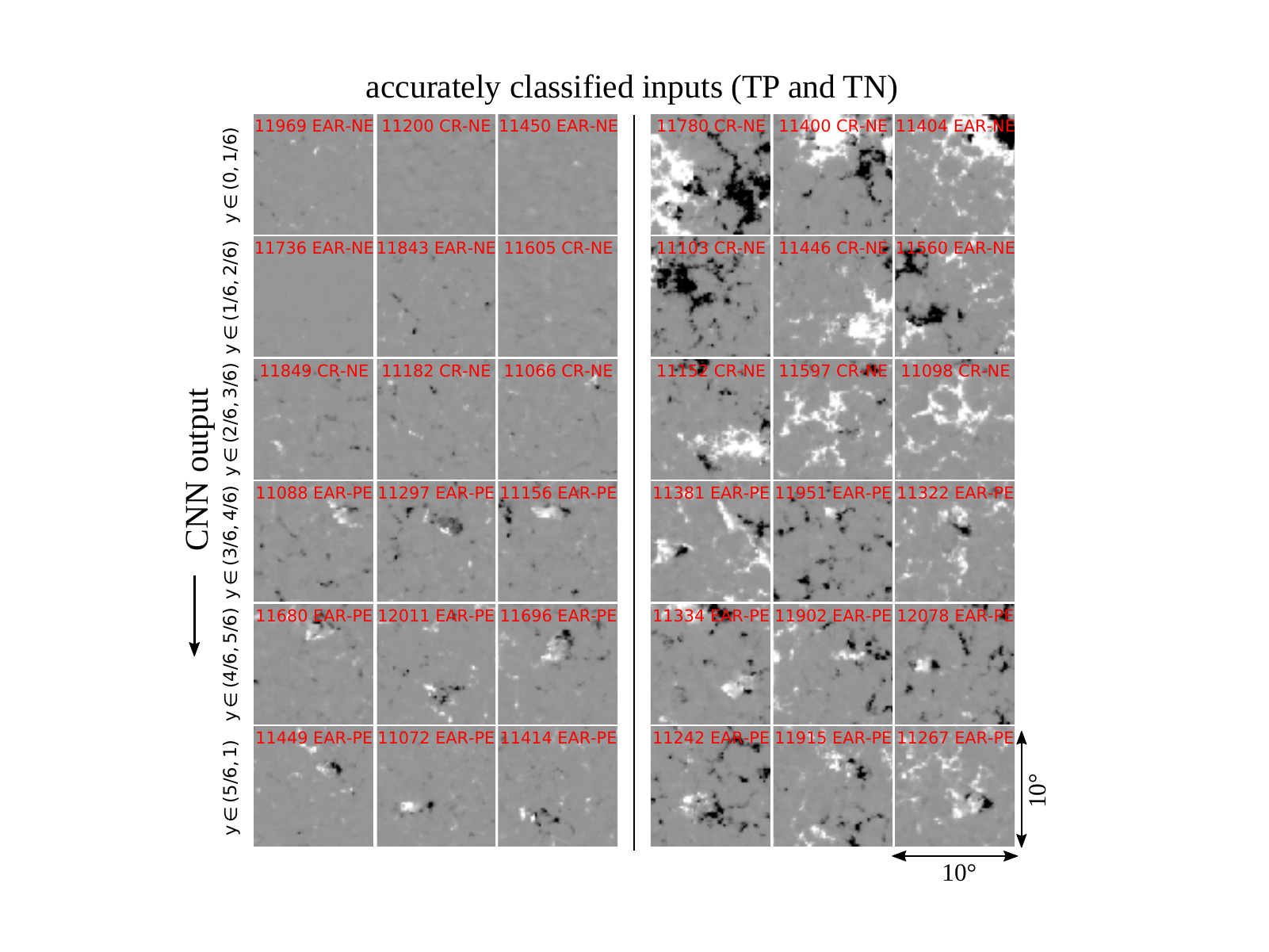}
\label{fig:trueInputs}
}
\subfloat
{
\includegraphics[width=0.50\textwidth,trim={2.3cm 0.8cm 2.7cm 0.8cm},clip]{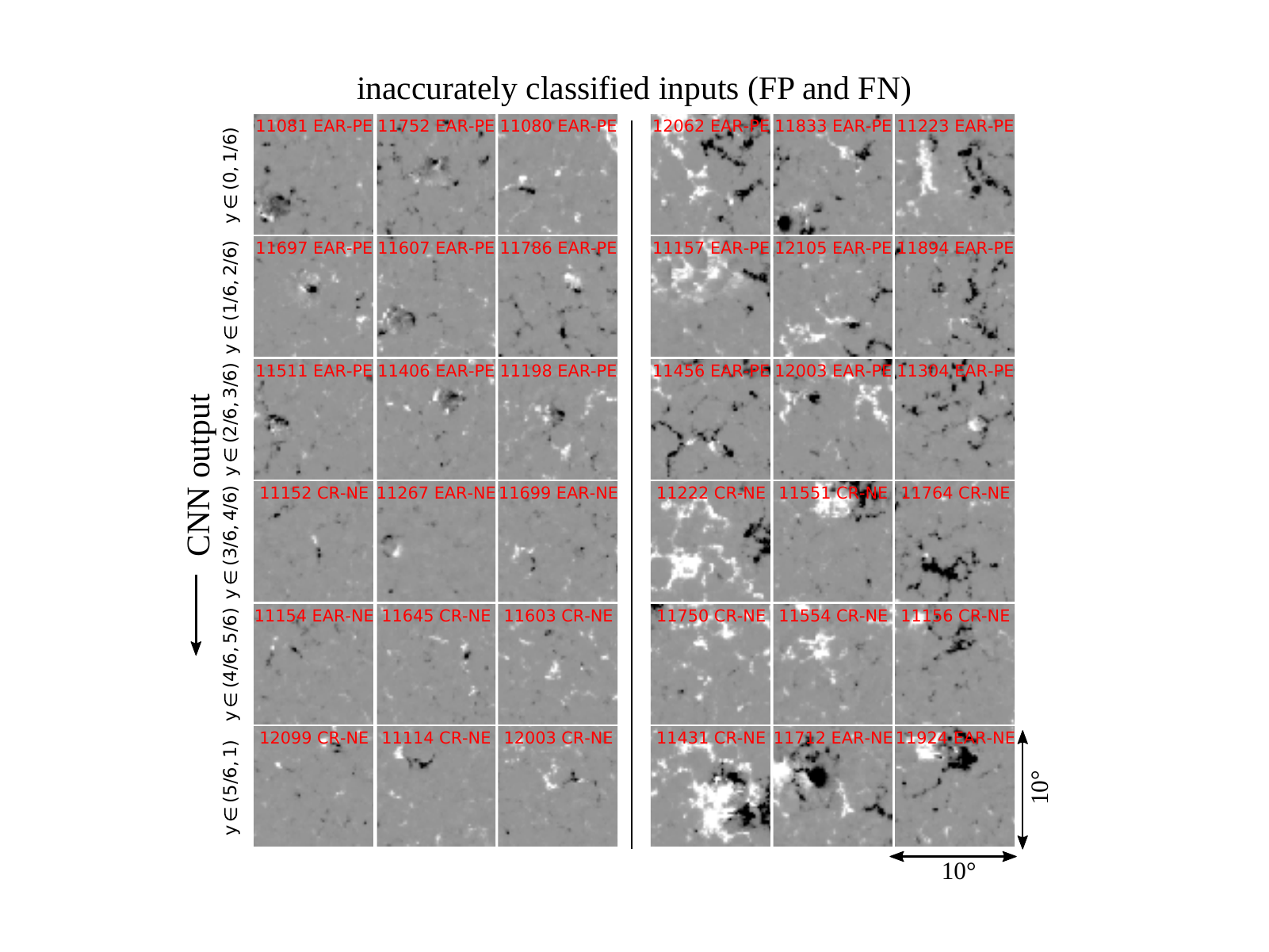}
\label{fig:falseInputs}
}

\caption{Categorizing input pre-emergence (PE) and non-emergence (NE) magnetogram samples (3.2~h before emergence) into bins corresponding to the CNN output values. For each bin, three samples with minimum total unsigned line-of-sight magnetic flux $\Flux$ ({\it left}) and three samples with maximum $\Flux$ ({\it right}) are shown. The bins are arranged such that the corresponding CNN output increases from {\it top} to {\it bottom}. {\it Left:} Magnetograms with accurately predicted class labels are shown. PE magnetograms classified accurately are true positives (TPs) and NE magnetograms classified accurately are true negatives (TNs). {\it Right:} Magnetograms with inaccurately predicted class labels are shown. PE magnetograms classified inaccurately are false negatives (FNs) and NE magnetograms classified inaccurately are false positives (FPs). The magnetogram plots are saturated at 150 G (white) and -150 G (black). For TNs ($y \sim 0$) magnetograms, $\Flux$ value can be very low as well as very high. For TPs ($y \sim 1$) magnetograms, $\Flux$ lies in an intermediate range.}
\label{fig;Inputs}
\end{figure*}

\begin{table}[b]
\centering
\begin{tabular}{cc}
\toprule
 depth of CNN & 5-fold cross-validation \\
(No. of hidden layers) & TSS ($\%$)\\
 \midrule
1 & 8.89 $\pm$ 17.57  \\
2 & 52.29 $\pm$ 7.88 \\
3 & 70.34 $\pm$ 3.64 \\
4 & 82.33 $\pm$ 2.87 \\
5 & 84.57 $\pm$ 6.40 \\
 \bottomrule
\end{tabular}
\caption{Mean 5-fold cross-validation {\it True Skill Statistics} (TSS) for classification of pre-emergence (PE) and non-emergence (NE) magnetogram segments at -3.2~h pre-emergence time obtained using the convolutional neural network (CNN, Figure~\ref{fig:CNN}) with increasing number of hidden convolutional layers. The $1\sigma$ error is quoted.}
\label{tab:TSSwithDepth}
\end{table}

\begin{table}[t]
\centering
\begin{tabular}{cc}
\toprule
time before  & 5-fold cross-validation \\
emergence (h) & TSS ($\%$)\\
 \midrule
-3.2 & 84.57 $\pm$ 6.40  \\
-8.5 & 61.97 $\pm$ 7.54 \\
 -13.9 & 48.68 $\pm$ 5.45 \\
 -19.2 & 40.80 $\pm$ 7.25 \\
-24.5 & 43.30 $\pm$ 6.93 \\
 \bottomrule
\end{tabular}
\caption{Mean 5-fold cross-validation {\it True Skill Statistics} (TSS) for classification of pre-emergence (PE) and non-emergence (NE) magnetograms at different pre-emergence times obtained using the convolutional neural network (CNN, Figure~\ref{fig:CNN}) with 5 hidden convolutional layers. The $1\sigma$ error is quoted.}
\label{tab:TSSwithTime}
\end{table}
We set up the CNN using Python's deep learning library {\it keras} \citep{chollet2015keras}. At the beginning of training, weights are initialized using the Glorot uniform initializer \citep{Glorot2010} and biases are initialized as zeros. The input magnetograms are standardized, i.e., the mean is subtracted and divided by the standard deviation. Note that the mean and standard deviation used for standardization is calculated over all PE and NE magnetogram samples. This limits the operational range of the pixel values, representing the magnetic field.

We obtain optimum training by setting learning rate $lr=1.0 \times 10^{-7}$ and batch size $N_{BS}=32$. We observe that the performance of CNN gradually improves with increasing depth (Table~\ref{tab:TSSwithDepth}). The max-pooling operation downsamples  feature maps obtained in each convolution layer by half. Therefore, we limit the depth of the CNN to 5 layers, to yield final layer feature maps (the final convolution layer outputs) of $6 \times 6$ pixel grid. Each pixel from the final layer feature map corresponds to approximately $20 \times 20~{\rm Mm}^2$ of the input PE and NE magnetograms, sufficiently large to incorporate the pre-emergence bipolar magnetic-field pattern \citep{Schunker2019}.   

Using the set of PE and NE magnetograms available at different pre-emergence times (see Table~\ref{tab:Dataset}), we train the CNN to yield maximum average 5-fold cross-validation TSS. We monitor training and validation losses at each iteration to ensure that they decrease monotonically and converge as training progresses, indicating that the network is well-trained and does not suffer from overfitting.  (Figure~\ref{fig:trainingLoss}).

The 5-fold cross-validation TSS obtained using the CNN trained with the set of magnetograms taken at different pre-emergence times, are listed in the Table~\ref{tab:TSSwithTime}. The TSS is $\sim 40\%$ at 24.5~h before emergence, comparable to the LBB survey results using the average unsigned radial magnetic field of EARs and CRs as discriminator. TSS increases as emergence approaches, yielding $\sim 85\%$ at 3.2~h before emergence, which is significantly higher compared to $\sim 53\%$ obtained in the LBB survey. 
\begin{figure*}[t]
\centering
\subfloat
{
\includegraphics[width=0.40\textwidth,trim={0cm 0cm 0cm 0cm},clip]{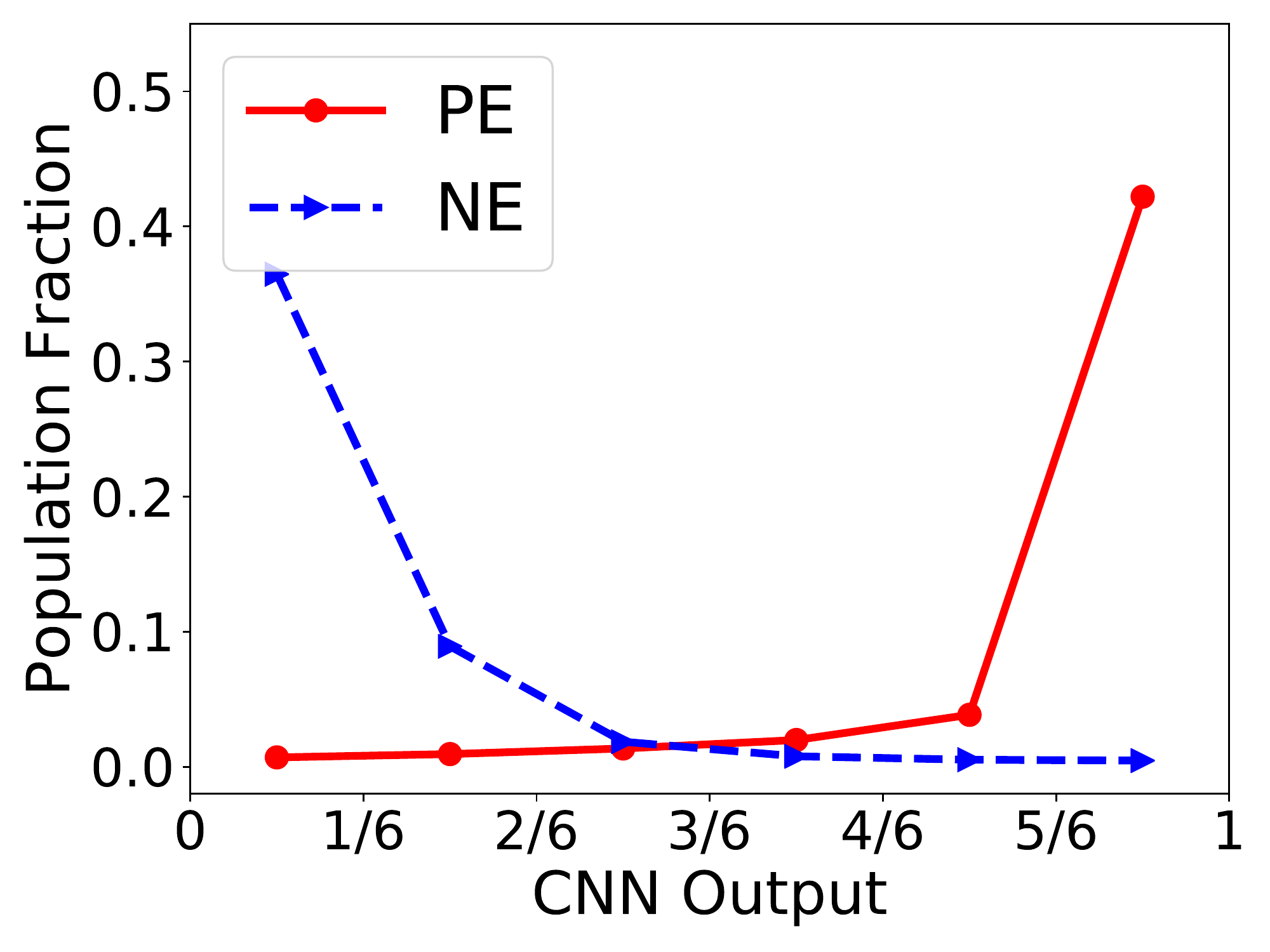}
\label{fig:predictedDistribution}
}
\subfloat{
\includegraphics[width=0.40\textwidth,trim={0cm 0cm 0cm 0cm},clip]{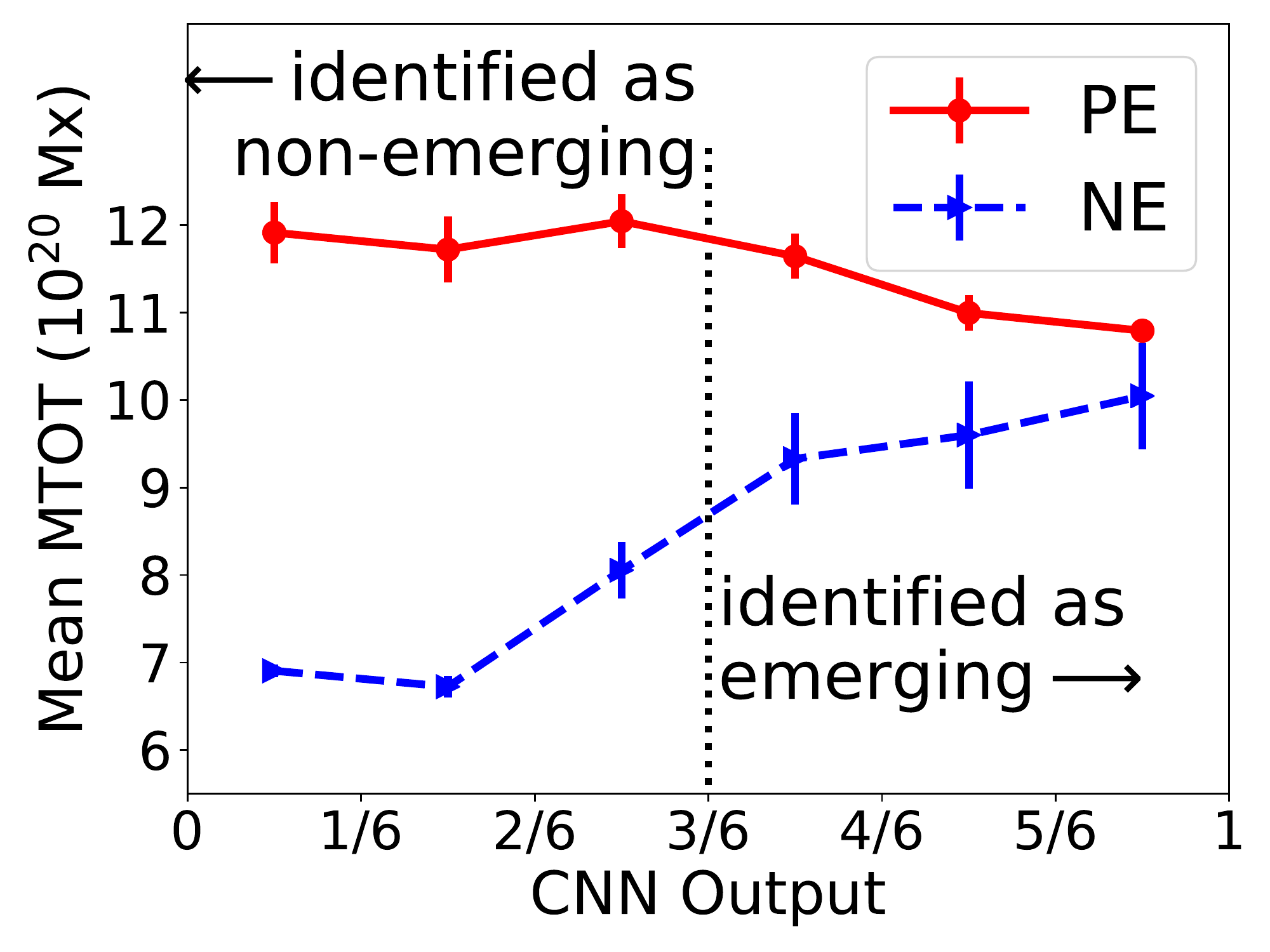}
\label{fig:binnedUnsignedFlux}
}
\caption{Statistical analysis of the total unsigned line-of-sight magnetic flux $\Flux$ of pre-emergence (PE) and non-emergence (NE) magnetograms taken at 3.2~h before emergence. The PE and NE samples are distributed in bins as per the corresponding CNN output. The bins of CNN output are centered at $y=\{1/12,3/12,5/12,7/12,9/12,11/12\}$ and bounded by $y \pm \Delta/2$ where $\Delta=1/6$. {\it Left:} The population density of PE and NE samples binned by the CNN output. For a significant majority of the PE samples $y \sim 1$ and for a significant majority of the NE samples $y \sim 0$. \textit{Right:} The average $\langle \Flux \rangle$ calculated over PE and NE samples from each bin of the CNN output. The PE and NE samples for which the CNN output $y \leq 0.5$ are labeled as non-emerging and $y > 0.5$ are labeled as emerging. The  ${\langle \Flux \rangle}_{\rm PE}$ decreases from a high value as the CNN output increases, whereas  ${\langle \Flux \rangle}_{\rm NE}$ increases from low value as the CNN output increases. Thus, NE samples with low CNN output $y\sim0$ have low values of  ${\langle \Flux \rangle}_{\rm NE}$ and PE samples with low CNN output $y\sim0$ have high values of ${\langle \Flux \rangle}_{\rm NE}$. For PE and NE samples with high CNN output $y \sim 1$, ${\langle \Flux \rangle}_{\rm PE}$ and ${\langle \Flux \rangle}_{\rm NE}$ values are in the intermediate range. The $3\sigma$ error bars are shown. 
} 
\label{fig:SA}
\end{figure*}

\subsection{Discriminant analysis of unsigned line-of-sight magnetic flux.}
The LBB survey used a non-parametric discriminant analysis of radial magnetic field $\overline{\vert {B}_{r} \vert}$, of $30\degree \times 30\degree$ EARs (CRs), averaged over 45.5~Mm about the emergence (center) location. The radial magnetic field was obtained from the line-of-sight {\it SOHO/MDI} data using a potential-field model. The present work is concerned with the classification of $10\degree \times 10\degree$ line-of-sight PE and NE magnetogram segments, obtained from EARs and CRs, using a CNN. We obtain a baseline for comparing the performance of the CNN using a non-parametric discriminant analysis, similar to the LBB survey, of the total unsigned line-of-sight magnetic flux $\Flux = \sum{\vert {B}_{\rm LOS} \vert~dA}$ of PE and NE magnetogram segments. $\Flux$ is a well-defined keyword in the SDO database \citep{Bobra2014}. The $\Flux$ measure used here is slightly different in that we do not require membership of a coherent magnetic field structure, but rather more simply all of the magnetic field within a specific area of $10\degree \times 10\degree$. 

For the discriminant analysis, we estimate probability density of $\Flux$ of the PE and NE magnetogram segments using the Epanechnikov kernel \citep{Silverman86,Barnes_2014}. We choose the kernel smoothing parameter value that is optimum for a normal distribution. We estimate the probability density for magnetograms in the training data. As shown in the left panel in Figure~\ref{fig:DA}, we find the discriminant boundaries 
where the estimated probability densities of $\Flux$ of PE and NE magnetograms are equal. These are used to classify magnetograms in the validation data and a TSS is obtained. Similar to the CNN, this process is performed for the five cross-validation sets.

From the left panel in Figure~\ref{fig:DA}, the distributions of $\Flux$ for PE and NE magnetogram segments are well-separated. For pre-emergence time -3.2~h, we find discriminant boundaries at $7.9\times10^{20}~\textrm{Mx}$ and $20.2\times10^{20}~\textrm{Mx}$  and at $0.2\times10^{20}~\textrm{Mx}$ and $2.7\times10^{20}~\textrm{Mx}$. The discriminant analysis yields cross-validation TSS of $65.8 \pm 4.5\%$. Repeating the analysis, we obtain TSS for classification of PE and NE magnetogram segments at different pre-emergence times. The right panel in Figure~\ref{fig:DA} shows that the CNN outperforms the discriminant analyses using the unsigned magnetic field ($\overline{\vert B_r \vert}$ or $\Flux$), most significantly for pre-emergence times -3.2~h and -8.5~h.

The CNN yields $\sim 20\%$ higher TSS for classification of PE and NE magnetograms compared to the baseline classification using discriminant analysis of $\Flux$. In the following, we attempt to interpret the CNN performance to understand the information learned for the classification of PE and NE magnetograms.

\subsection{The mapping between the unsigned line-of-sight magnetic flux and the CNN output.}
Given an input PE or NE magnetogram segment, the trained CNN outputs a value between 0 and 1. The predicted binary class labels 0 and 1 are obtained by thresholding the output at 0.5. The original CNN output can be interpreted as the probability of the input magnetogram segment showing emergence after a certain time. Thus, the trained CNN maps surface magnetic field (input) to the probability of emergence (output). Patterns common to the magnetograms for which the CNN output is $y \sim 0$ are weakly correlated with emergence and patterns common to the magnetograms for which the CNN output $y \sim 1$ are strongly correlated with emergence. 

We visually inspect patterns common to the PE and NE magnetogram segments by arranging according to corresponding CNN output into six bins centered at CNN output values $y=\{1/12,3/12,5/12,7/12,9/12,\allowbreak 11/12\}$, bounded by $y \pm \Delta/2$ where bin-width $\Delta = 1/6$. The left panel of Figure~\ref{fig;Inputs} shows representative magnetograms (taken at -3.2~h pre-emergence) with accurate predicted class labels from the six CNN output bins. We see that the CNN output is low $y \sim 0$ (top row) for the magnetograms with very low as well as very high value of $\Flux$. With the CNN output progressively increasing for the magnetograms in the subsequent bins, the corresponding value of $\Flux$ either progressively increases from the low value or progressively decreases from the high value. The magnetograms that yield $y \sim 1$ (bottom row) are the  magnetograms with $\Flux$ value within an intermediate range. The high value of  ${\rm TSS} \sim 85\%$, for classification of PE and NE magnetogram segments using the CNN, implies that the distribution of $\Flux$ of magnetograms within this intermediate range closely matches the distribution of $\Flux$ for the PE samples.

The right panel of Figure~\ref{fig;Inputs} similarly shows magnetogram samples with inaccurately predicted class labels arranged in the six bins of CNN output. Overall, there are significantly fewer such magnetogram segments compared to the number of magnetograms with accurately predicted class labels (Figure~\ref{fig:SA}). The PE magnetogram segments that yield $y \sim 0$ are relatively unclean emergences i.e. emerging in the vicinity of strong pre-existing magnetic field. The NE magnetogram segments that yield $y \sim 1$ show distinct bipolar magnetic-field structure akin to the pre-emergence active region (Figure~\ref{fig:EAR-CR_Averaged}). Few NE magnetogram segments with moderately high $\Flux$ also yield high CNN output $y \sim 1$.

Figure~\ref{fig;Inputs} clearly shows that there is a wide range of $\Flux$ values for many of the CNN output bins. Therefore, we analyse $\Flux$ of PE and NE samples from each category of the CNN output bins. We only consider a sub-population of PE and NE magnetograms between $\Flux$ of $4 \times 10^{20}~{\rm Mx}$ and $16 \times 10^{20}~{\rm Mx}$ (Figure~\ref{fig:subpopulation}) and discard a small number of PE and NE samples with very high and very low values of $\Flux$ (see Appendix~\ref{app:subpop}). The left panel of Figure~\ref{fig:SA} shows the distribution of PE and NE magnetograms binned according to the corresponding CNN output. The PE distribution clusters at $y \sim 1$ and the NE distribution clusters at $y \sim 0$.

We calculate the average value ${\langle \Flux \rangle}_{\rm PE/NE}^{\rm bin} = (1/N_{\rm PE/NE}^{\rm bin}) \sum_{N_{\rm PE/NE}^{\rm bin}} \Flux$, over PE and NE samples ($N_{\rm PE}^{\rm bin}$ and $N_{\rm NE}^{\rm bin}$) from each ${\rm bin}$ of the CNN output. The right panel of Figure~\ref{fig:SA} shows the variation of the mean value of ${\langle \Flux \rangle}_{\rm PE/NE}$ with CNN output. PE segments with low CNN output $y \sim 0.1$ have high ${\langle \Flux \rangle}_{\rm PE} \sim 12\times10^{20}~\textrm{Mx}$ and NE segments with low CNN output $y \sim 0.1$ have low ${\langle \Flux \rangle}_{\rm NE} \sim 7 \times 10^{20}~\textrm{Mx}$. As the CNN output increases, ${\langle \Flux \rangle}_{\rm PE}$ decreases and ${\langle \Flux \rangle}_{\rm NE}$ increases. The PE and NE samples that produce output $y > 0.5$ fall within an intermediate range of ${\langle \Flux \rangle}$ between $9\times10^{20}~\textrm{Mx} - 11\times10^{20}~\textrm{Mx}$. Thus, $\Flux$ is an important factor contributing to the CNN performance such that magnetograms with the CNN output $y \geq 0.5$ have average $\Flux$ value within an intermediate range (between $9\times10^{20}~\textrm{Mx} - 11\times10^{20}~\textrm{Mx}$). As shown in the left panel of Figure~\ref{fig:DA}, the population fraction of the PE magnetograms peaks within the $\Flux$ range of $9\times10^{20}~\textrm{Mx} - 11\times10^{20}~\textrm{Mx}$. Therefore, $\langle\Flux\rangle$ values in the right panel of Figure~\ref{fig:SA} are influenced by the original distribution of PE and NE magnetograms with respect to $\Flux$. Explicit analysis of CNN output as a function of $\Flux$ shows that there are other factors contributing to the CNN performance (see Appendix~\ref{app:MTOT}).

\subsection{Mapping between unsigned line-of-sight magnetic flux and the convolutional filter outputs.}
To further understand the performance of the CNN, we analyse the contribution of different convolutional filters towards the CNN output. We focus on the final convolutional layer of the CNN (Figure~\ref{fig:CNN}), which is directly connected to the output layer. The final layer consists of 256 convolutional filters (Figure~\ref{fig:ConvFilter}) and may learn up to 256 feature maps, i.e., surface magnetic field patterns correlated with emergence (a significant number of which may be redundant). It is challenging to analyse all 256 filters in detail and understand the contribution of each filter. It is however possible to systematically reduce the number of filters in the final layer, during training, to only few filters which are easier to interpret. This process is known as network pruning \citep{Cun90,Han2015,Frankle2019,Qin2018}. 

We develop a network pruning algorithm to iteratively reduce the final convolutional layer filters from $N=256$ to $N=4$.  The pruning algorithm is based on identifying top $N/4$ filters maximally correlated and top $N/4$ filters maximally anti-correlated with surface magnetic-field patterns associated with emergence and removing the remaining relatively neutral $N/2$ filters (Appendix~\ref{app:pruning}). The four filters in the final convolutional layer at the end of the training are the dominant four filters contributing towards the CNN output, two corresponding each to surface magnetic-field patterns correlated and anti-correlated with emergence.

\begin{figure}[t]
\centering
\includegraphics[width=0.48\textwidth,trim={0cm 0cm 0cm 0cm},clip]{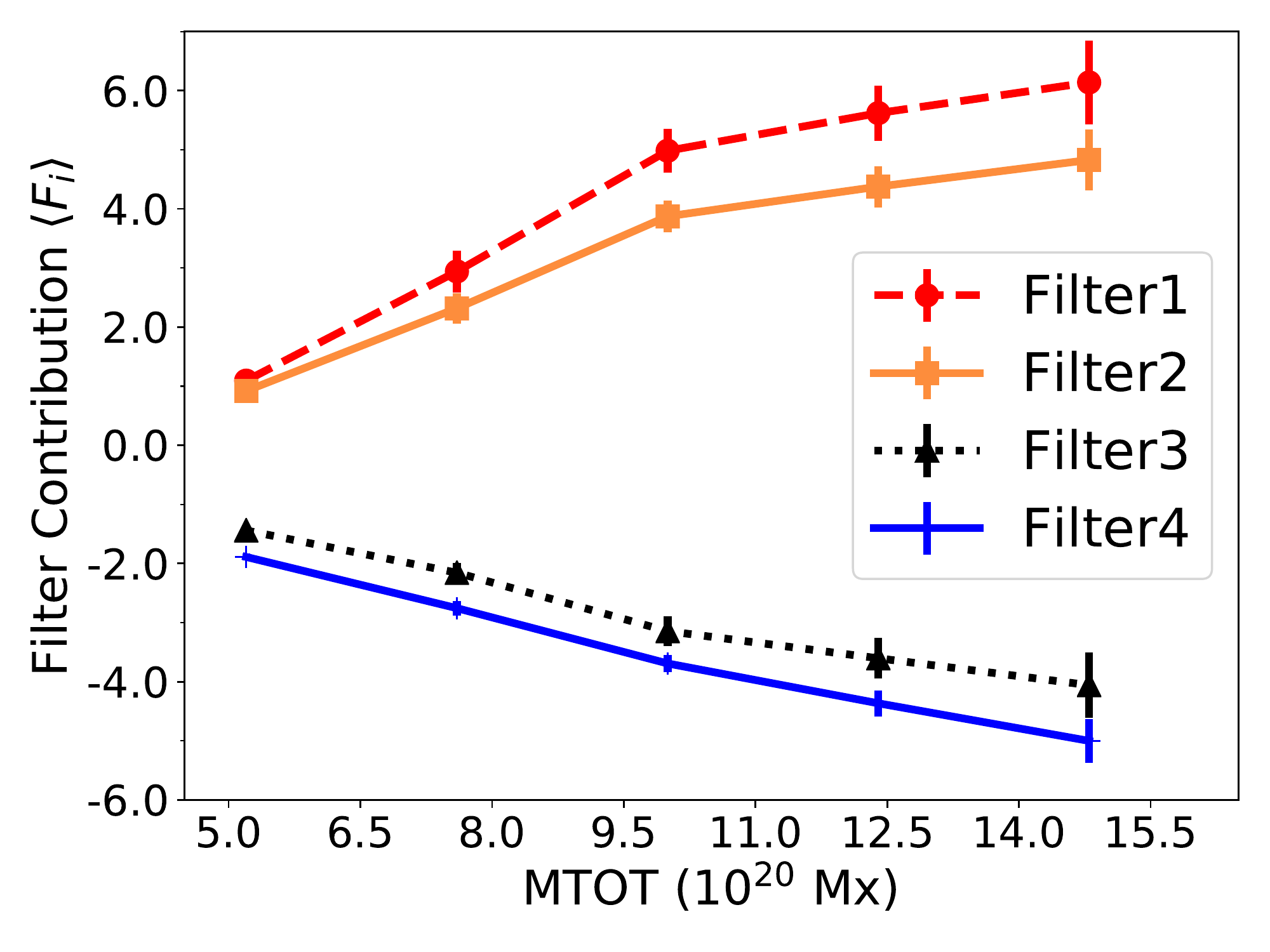}
\caption{Filter contributions (maximum positive value times the filter weight, Eq.~\ref{eq:filterContri})  of the final convolutional layer in the CNN trained with pruning algorithm (Appendix~\ref{app:pruning}). The PE and NE samples are categorized in five bins according to the total unsigned line-of-sight magnetic flux ($\Flux$). Average filter contribution ${\langle F_{i} \rangle}^{\rm bin}$ corresponding to PE and NE samples from each bin and each filter $i$ is plotted.  Contributions from filters 1 and 2 increase with growing $\Flux$, whereas filters 3 and 4 show decreasing contribution with increasing $\Flux$. Thus, the CNN incorporates filters that respond positively as well as negatively to increasing $\Flux$. $20\sigma$ error bars are shown.}
\label{fig:binnedsFilterActs}
\end{figure}
\begin{figure*}[t]
\centering
\subfloat
{
\includegraphics[width=0.50\textwidth,trim={2.4cm 0.9cm 3.2cm 1.3cm},clip]{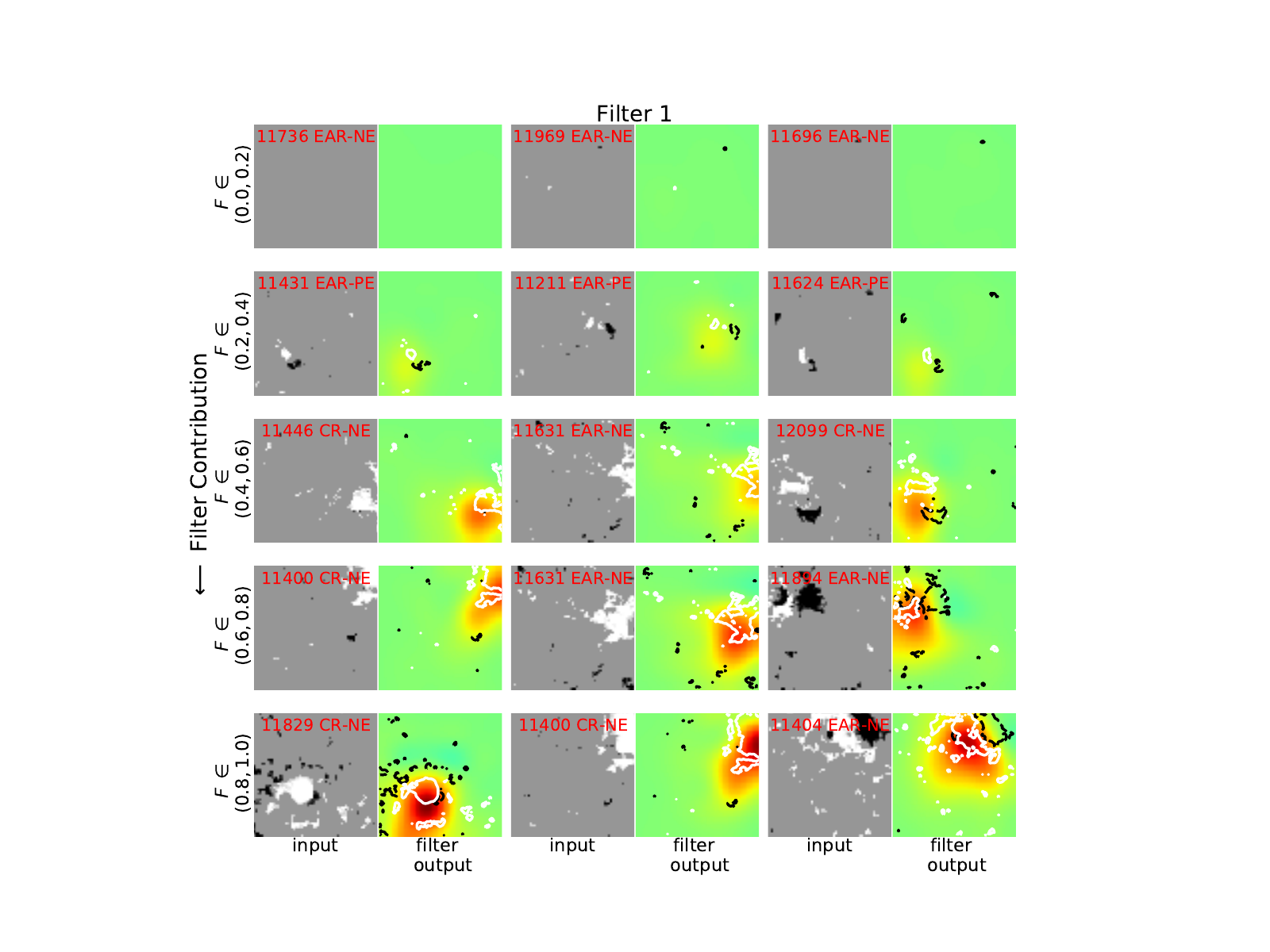}
\label{fig:filterActVis0}
}
\subfloat
{
\includegraphics[width=0.50\textwidth,trim={2.4cm 0.9cm 3.2cm 1.3cm},clip]{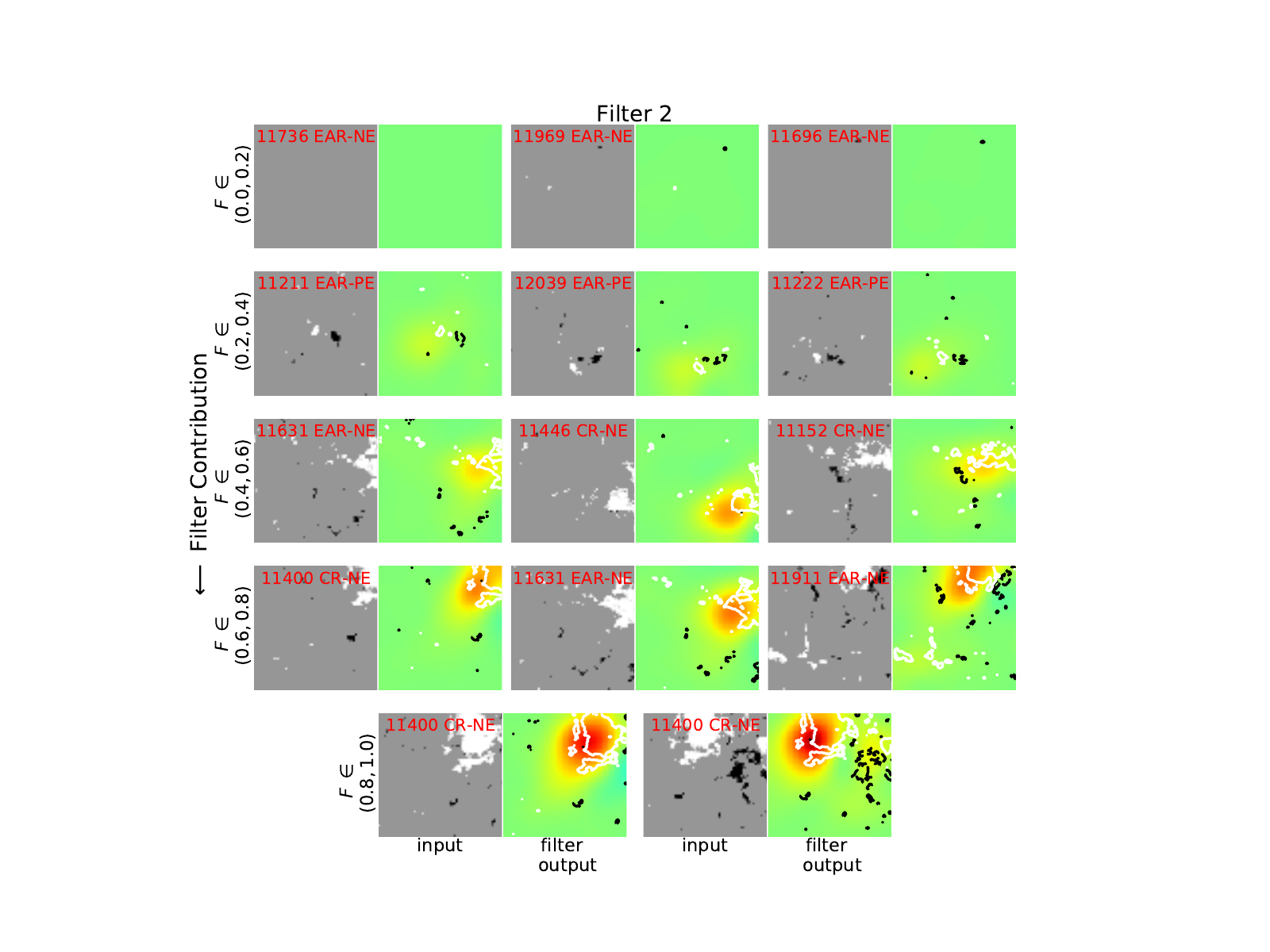}
\label{fig:filterActVis1}
}\\
\subfloat
{
\includegraphics[width=0.50\textwidth,trim={2.4cm 0.9cm 3.2cm 1.3cm},clip]{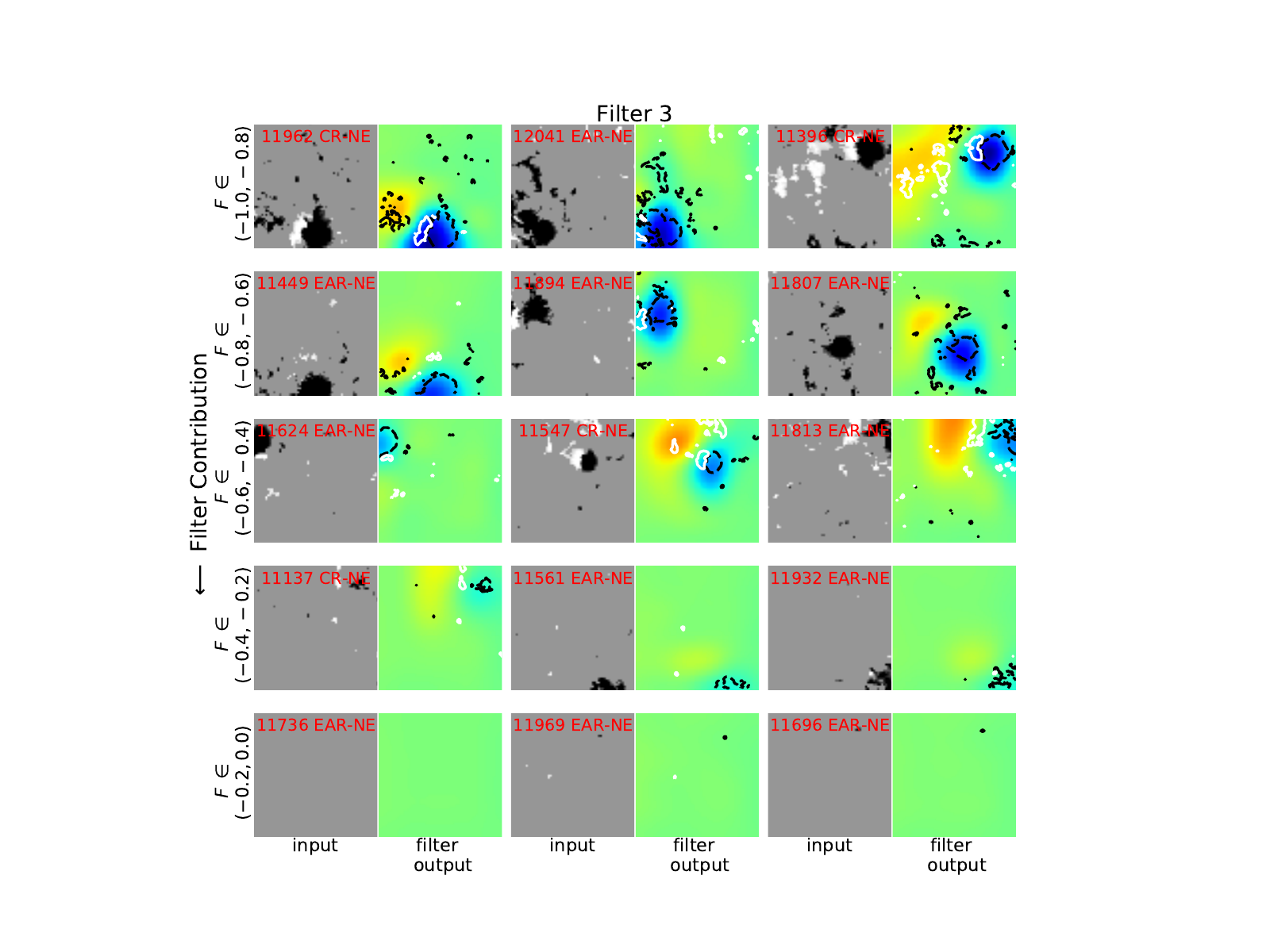}
\label{fig:filterActVis2}
}
\subfloat
{
\includegraphics[width=0.50\textwidth,trim={2.4cm 0.9cm 3.2cm 1.3cm},clip]{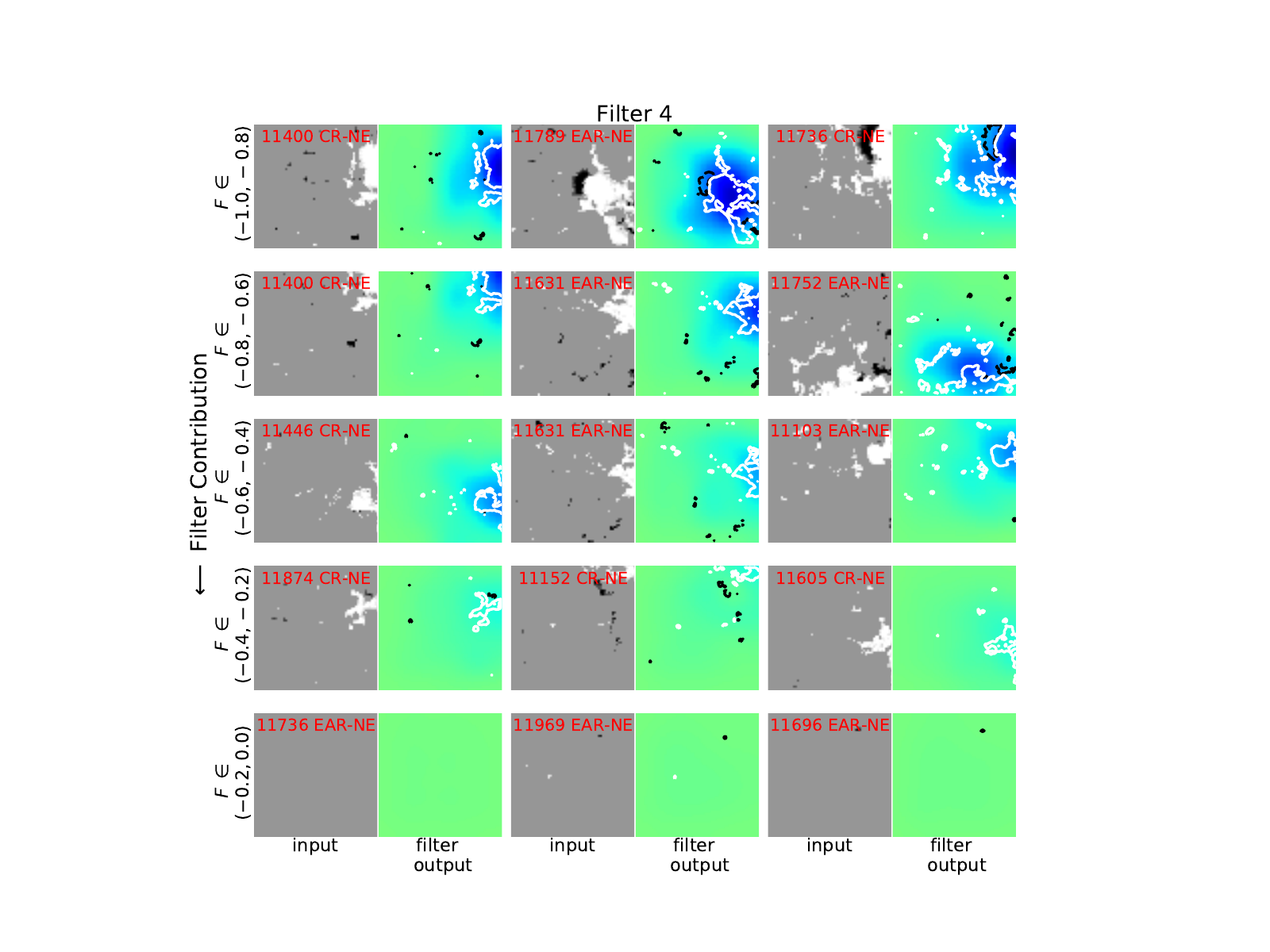}
\label{fig:filterActVis3}
}
\caption{Outputs of the filters in the final convolutional layer of the CNN for representative PE and NE samples. The CNN is trained using network pruning to retain the four most important filters in the final convolutional layer (Appendix~\ref{app:pruning}). The PE and NE samples are binned by the normalized filter contribution $F_{i}$ (maximum positive value times the filter weight, Eq.~\ref{eq:filterContri}), arranged from top to bottom in increasing order of the filter output. For each bin, three PE or NE samples with minimum total unsigned line-of-sight magnetic flux ($\Flux$) are shown (except the {\it bottom} bin for filter 1 in panel b, where only two are available). The input PE and NE magnetograms are plotted with magnetic field saturated at 150~G (white) and -150~G (black) and are clipped below an absolute value of 60~G. Filter outputs are saturated at 1 (red) and -1 (blue). 100~G (white) and -100~G (black) contours from the magnetograms are shown on top of the filter outputs for clarity. Filters 1 and 2 produce positive contributions $F_{i}$, while filters 3 and 4 produce negative contributions. For filters 1 and 2, the minimum $\Flux$ of PE and NE samples increases as the filter contribution increases and conversely for filters 3 and 4. For filters 1 and 2, the output is maximum in the region of large magnetic flux in the input. The filter 4 output also scales with the local magnetic flux in the input, but with a negative sign. For filter 3, the output is positive and negative corresponding to magnetic-flux regions.}
\label{fig:filterActVis}
\end{figure*}
\begin{figure*}[t]
\centering
\includegraphics[width=\textwidth,trim={0.0cm 4.8cm 0.0cm 4.5cm},clip]{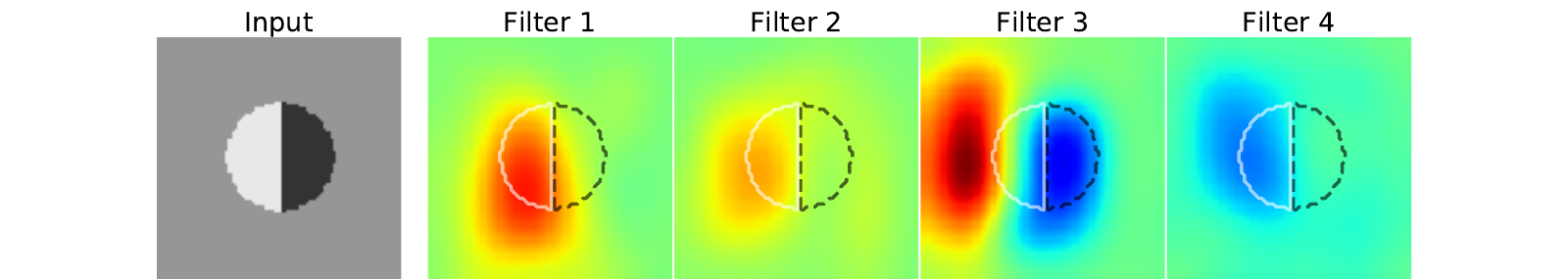}
\caption{Outputs of filters in the final convolutional layer of the pruned CNN for a synthetic bipole of 100~G uniform field and radius 25~Mm (Figure~\ref{fig:synBipole}). Filter outputs are saturated at 1 (red) and -1 (blue). 100~G (white) and -100~G (black) contours from the synthetic bipole are shown on top of the filter outputs. Filter outputs conform to the edges of the positive and negative polarity regions.} 
\label{fig:synFilterVis}
\end{figure*}

\subsubsection{Outputs of the filters in the final convolutional layer.}
We analyse the response of the four convolutional filters in the final layer of the pruned CNN with respect to $\Flux$. We divide the sub-population of PE and NE samples (Appendix~\ref{app:subpop}) in five bins of equal widths. For PE and NE samples in each bin, we calculate filter contribution $F_{i}$ (maximum positive value multiplied by the filter weight, Eq.~\ref{eq:filterContri}) for each of the four filters in the pruned CNN. We obtain the average value of filter contribution ${\langle F_{i}\rangle}^{\rm bin}$ for each bin, for each filter $i$. Figure~\ref{fig:binnedsFilterActs}, shows filters 1 and 2, which correspond to surface magnetic-field patterns that are correlated with emergence, ${\langle F_{i}\rangle}^{\rm bin}$ increases with increasing $\Flux$. Similarly, for filters 3 and 4, which correspond to surface magnetic-field patterns that are anti-correlated with emergence, ${\langle F_{i}\rangle}^{\rm bin}$ decreases with increasing $\Flux$. Thus, the trained CNN incorporates filters which respond both positively as well as negatively to $\Flux$ of the PE and NE magnetograms. The magnitude of filter contributions ${\langle F_{i}\rangle}^{\rm bin}$ increases monotonically with increasing $\Flux$.

To further illustrate the operation of the convolutional filters in the pruned CNN, we visibly inspect the output of each filter for a few representative PE and NE samples. Figure~\ref{fig:filterActVis} shows select examples of outputs of each convolutional filter, binned by the normalized contributions (Eq.~\ref{eq:filterContri}) from the filter. Filters 1 and 2 contribute positively towards the CNN output. Also the (minimum) $\Flux$ of the magnetogram samples increases with filter contribution.  We see that the filter output is maximum  for the prominent magnetic-flux regions in the input magnetograms, yielding positive filter output. Filters 3 and 4 contribute negatively to machine prediction. Also, the (minimum) $\Flux$ of the magnetogram samples decreases as filter contribution increases. Filter 3 produces positive and negative outputs corresponding to magnetic-flux areas. Filter 4 produces negative output corresponding to magnetic-flux areas of the input magnetograms. Information about the polarity of the magnetic flux areas may not be necessary to train a CNN
that performs comparably to one trained on data including the polarity of the field (see Appendix~\ref{app:ABSData}). Also, these filter outputs spatially conform to the edges of the magnetic regions (see Figure~\ref{fig:synFilterVis}). 
\begin{figure}[t]
\centering
\subfloat
{
\includegraphics[width=0.22\textwidth,trim={0.7cm 0.4cm 0.9cm 0.9cm},clip]{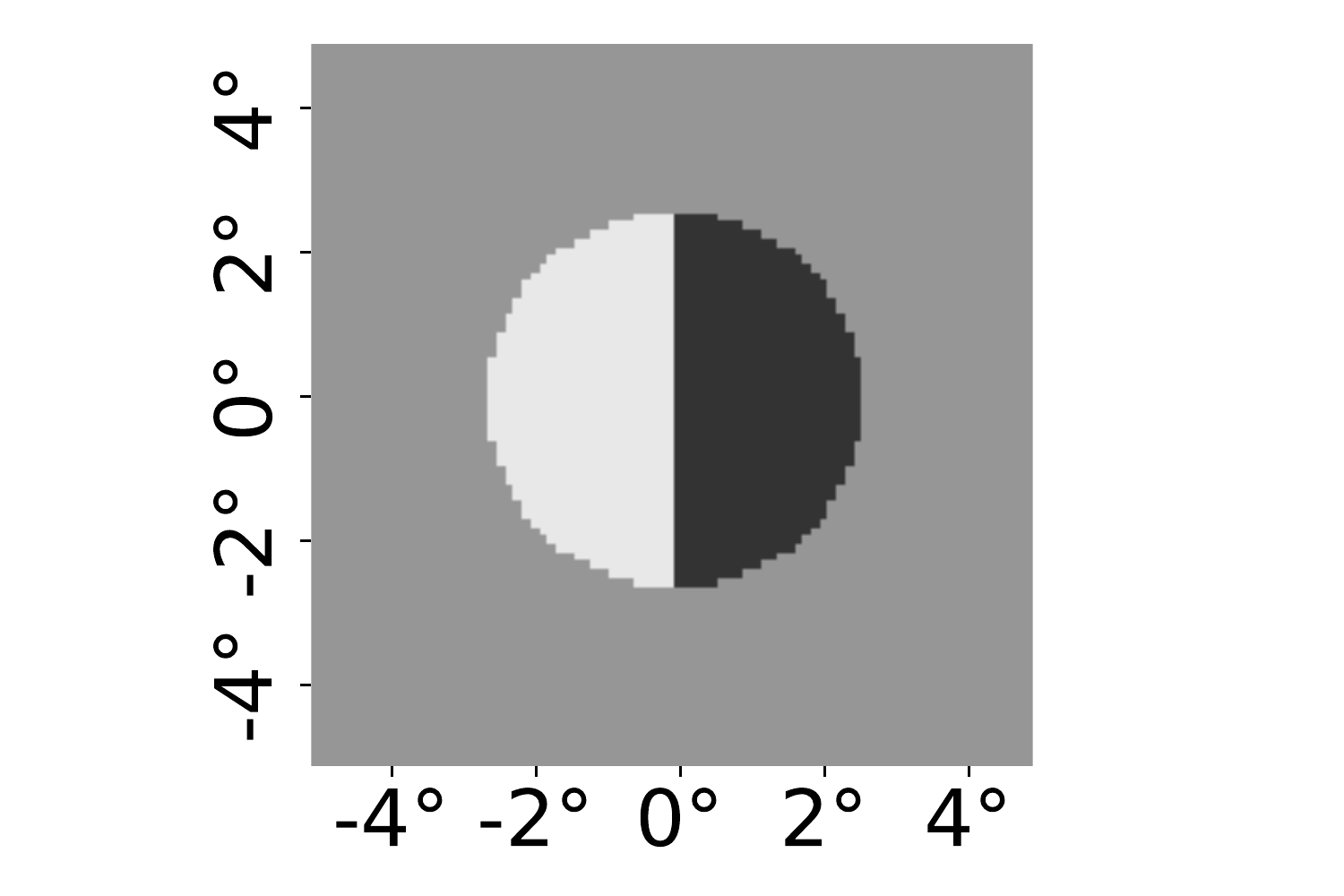}
\label{fig:synBipole1}
}
\subfloat
{
\includegraphics[width=0.24\textwidth,trim={0.40cm 0.25cm 0.4cm 0.4cm},clip]{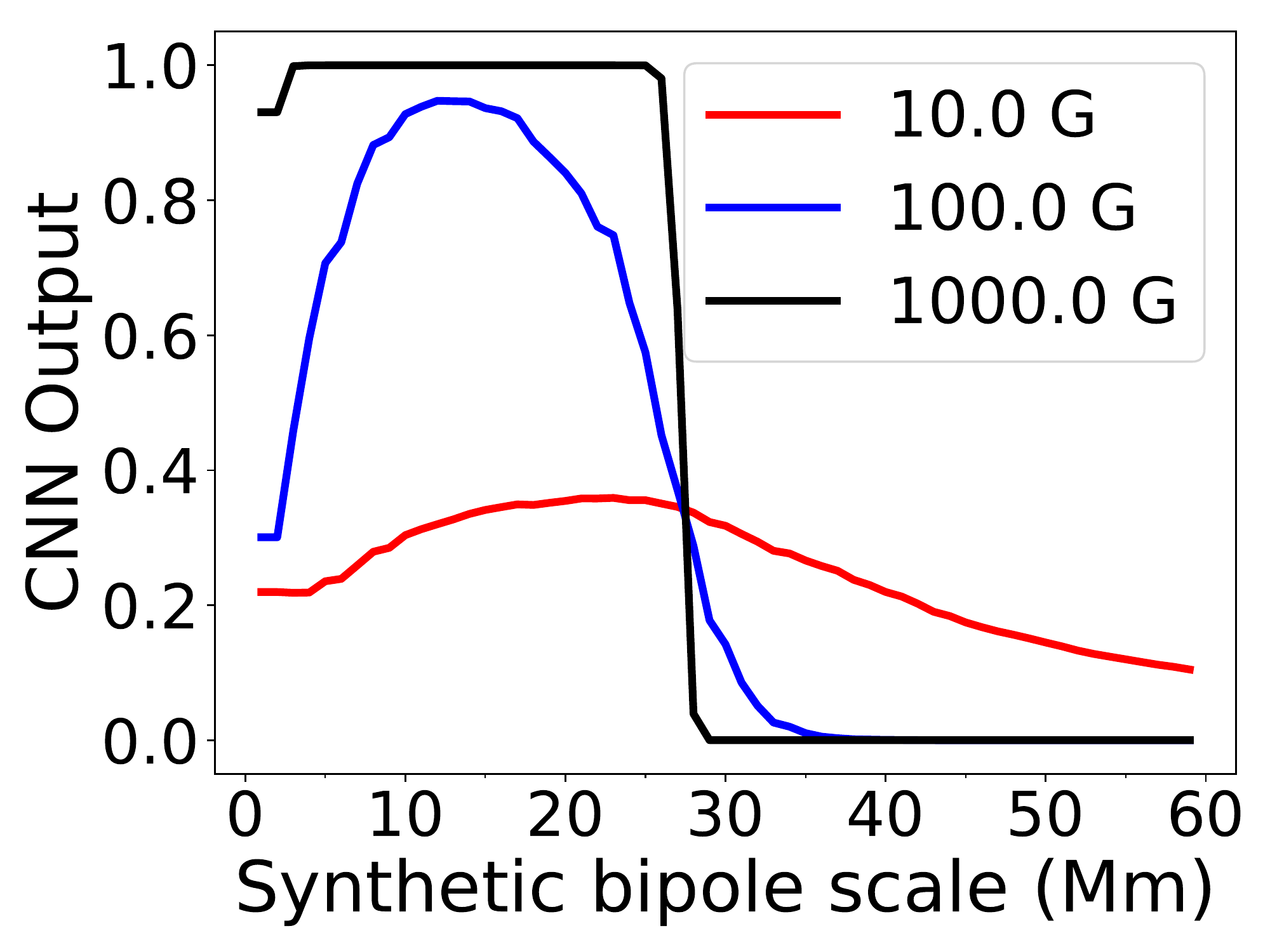}
\label{fig:synBipole2}
}
\caption{{\it Left:} Circular synthetic bipole of uniform magnetic field strength used for probing the trained CNN. {\it Right:} Output of the pruned CNN as a function of (uniform) intensity of the magnetic field and synthetic bipole length-scale (radius). For 100~G and 1000~G, there is a characteristic length-scale for the CNN beyond which output sharply drops to $0$. } 
\label{fig:synBipole}
\end{figure}
\begin{figure}[b]
\centering
\includegraphics[width=0.30\textwidth,trim={0cm 0cm 0cm 0cm},clip]{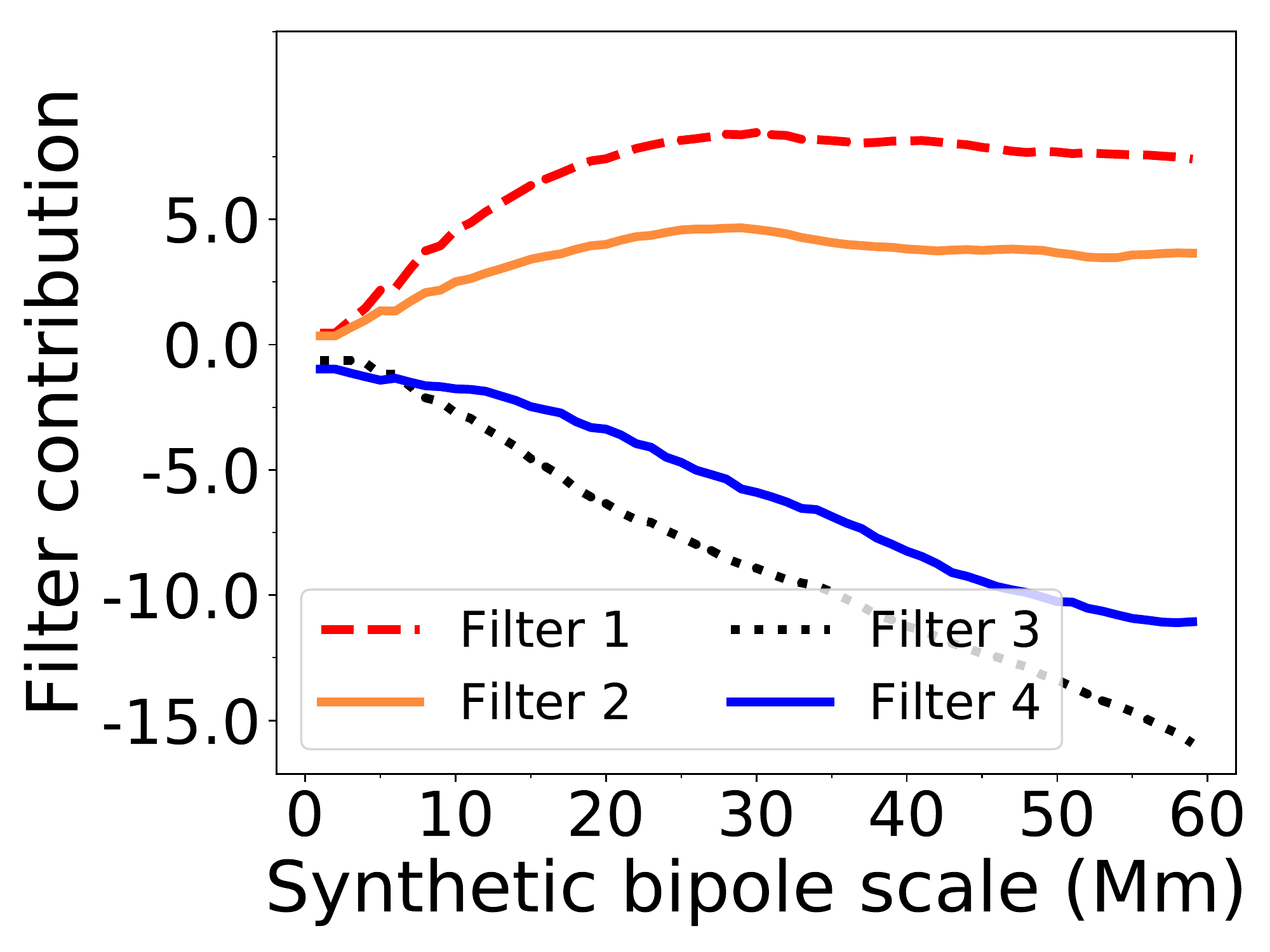}
\caption{Filter contributions $F_{i}$ (maximum positive value times the filter weight, Eq.~\ref{eq:filterContri}) of the pruned CNN as a function of the synthetic bipole (Figure~\ref{fig:synBipole}) length-scale (radius) for 100~G uniform magnetic field.} 
\label{fig:synFilters}
\end{figure}
\begin{figure}[b]

\centering
\subfloat
{
\includegraphics[width=0.24\textwidth,trim={0.0cm 0.0cm 0.0cm 0.0cm},clip]{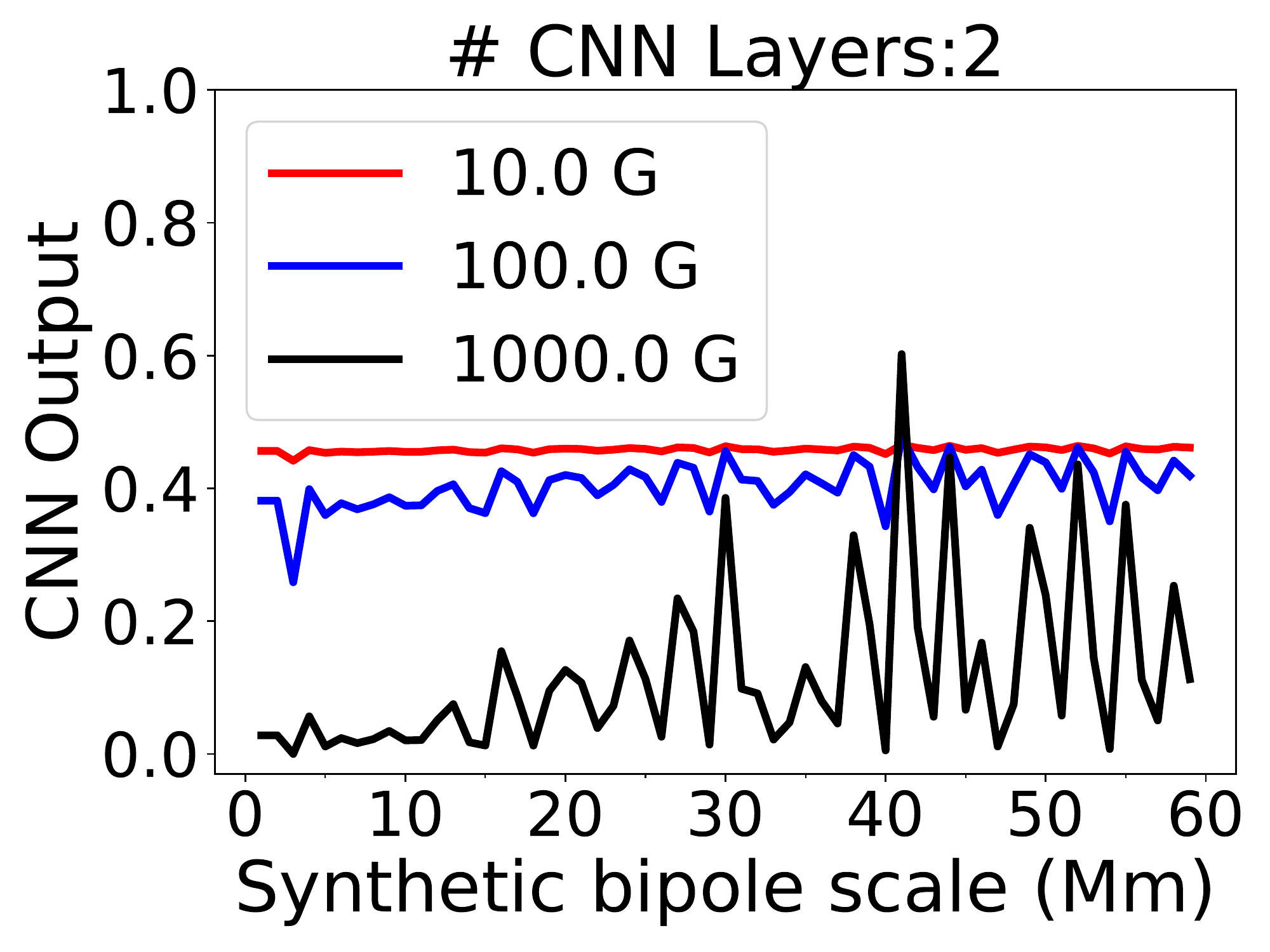}
\label{fig:synDepth2}
}
\subfloat
{
\includegraphics[width=0.24\textwidth,trim={0.0cm 0.0cm 0.0cm 0.0cm},clip]{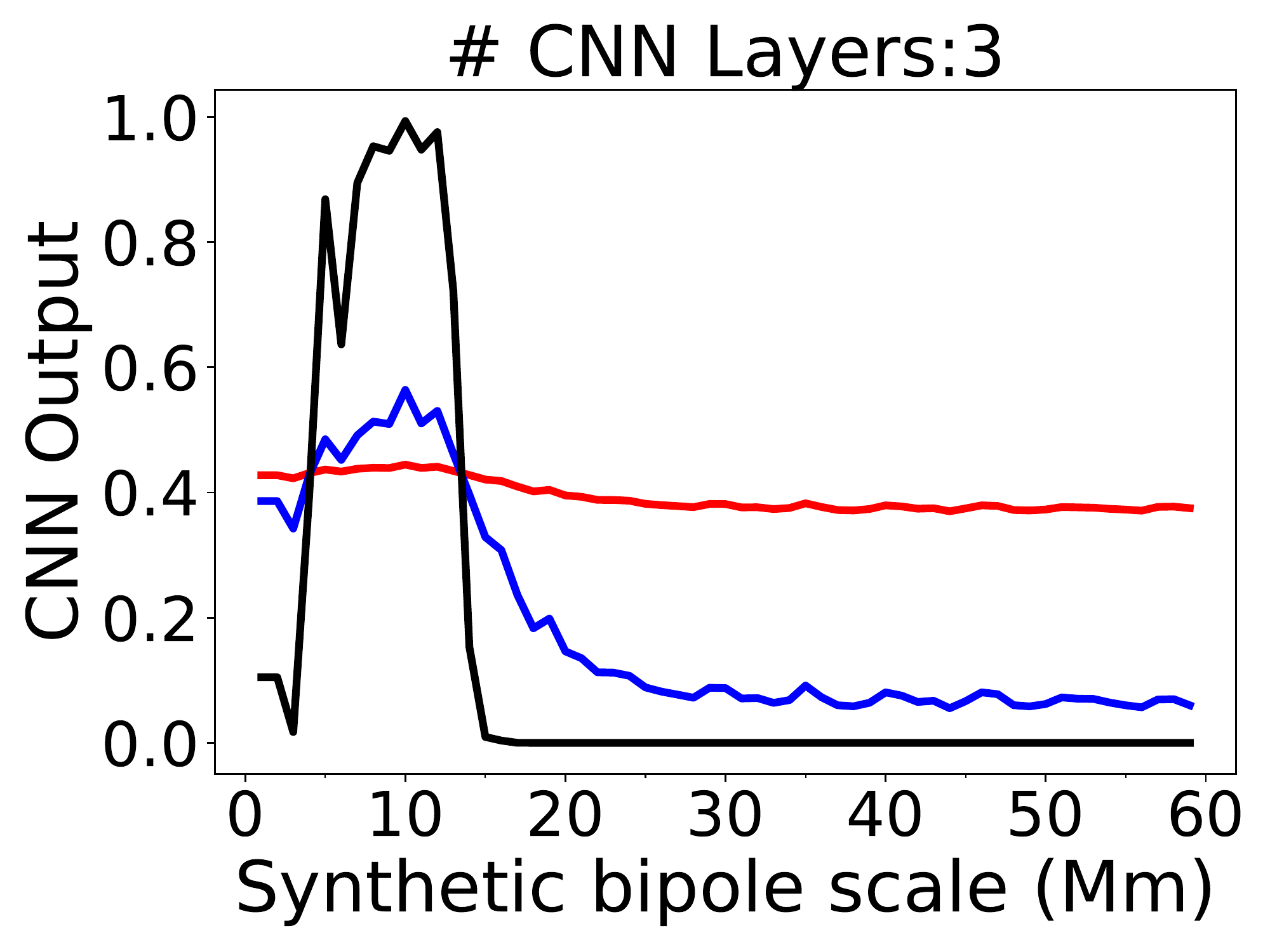}
\label{fig:synDepth3}
}\\
\subfloat
{
\includegraphics[width=0.24\textwidth,trim={0.0cm 0.0cm 0.0cm 0.0cm},clip]{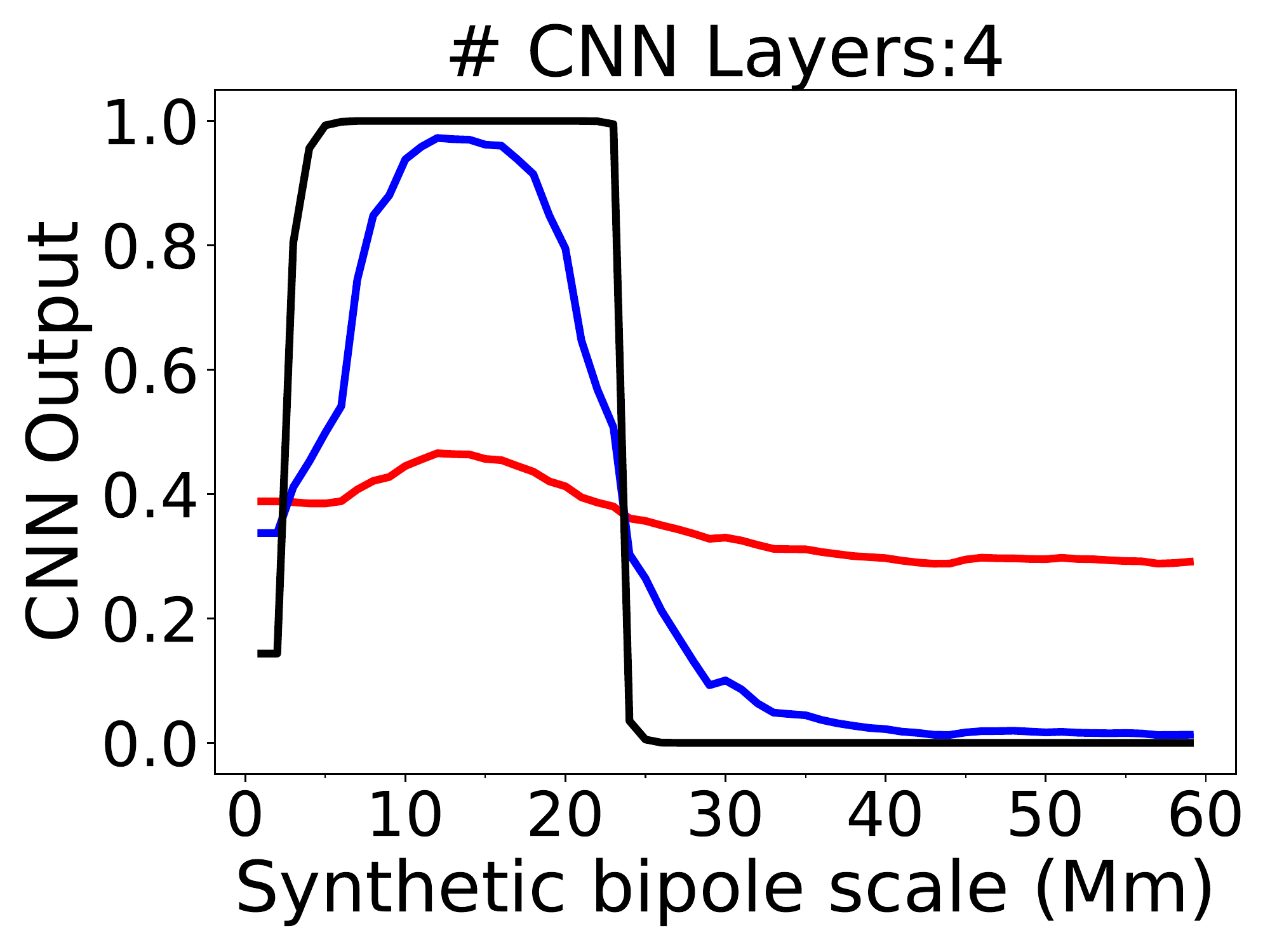}
\label{fig:synDepth4}
}
\subfloat
{
\includegraphics[width=0.24\textwidth,trim={0.0cm 0.0cm 0.0cm 0.0cm},clip]{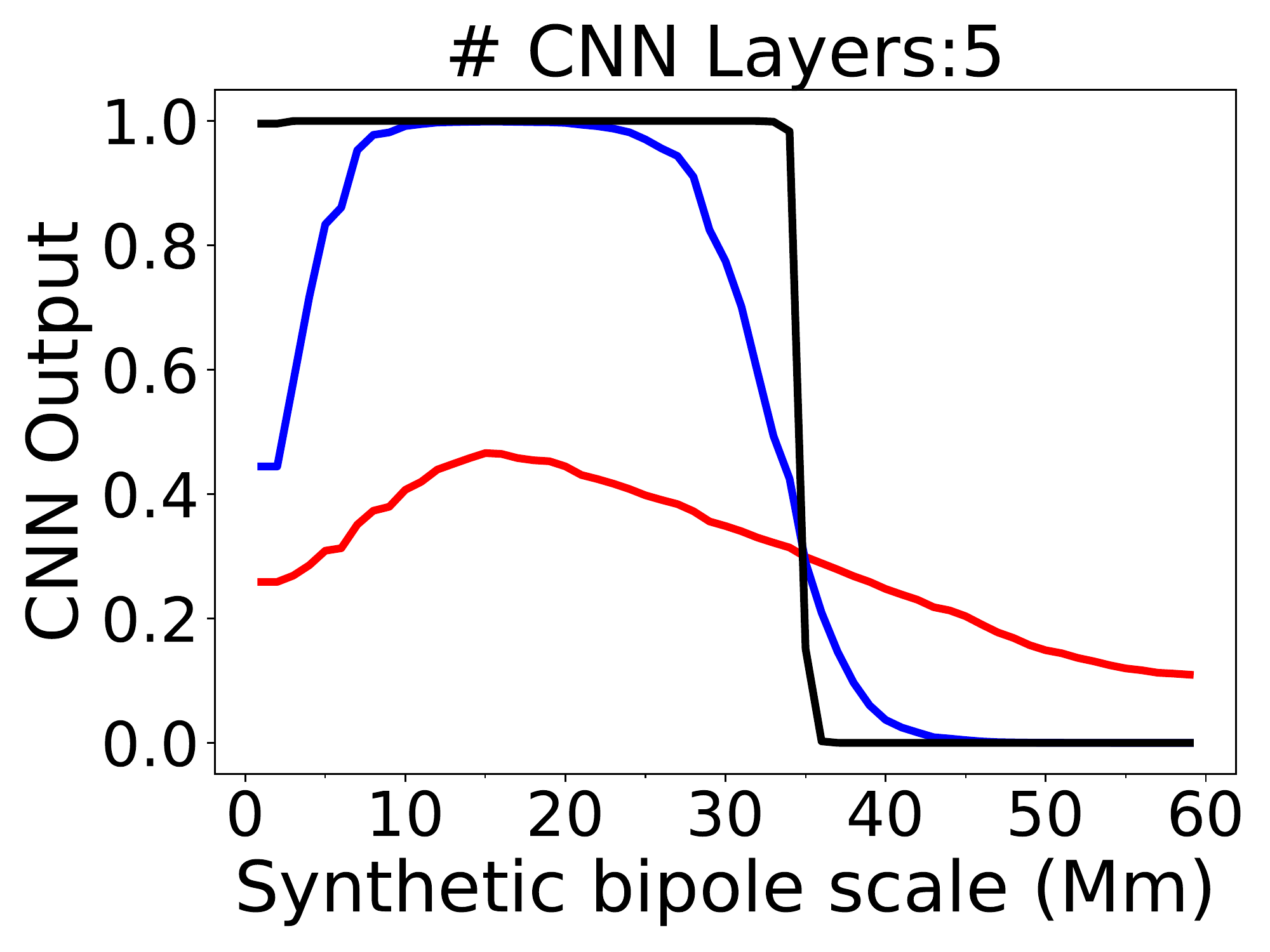}
\label{fig:synDepth5}
}
\caption{The CNN output for synthetic bipoles for CNNs with increasing number of convolutional layers. The correlation between the CNN output and length-scale of the synthetic bipoles exists for CNN with more than two convolutional layers. The characteristic length-scale, beyond which the CNN output sharply drops, increases with increasing number of CNN layers.} 
\label{fig:synDepth}
\end{figure}
\begin{figure*}[t]
\centering
\subfloat
{
\includegraphics[width=0.50\textwidth,trim={1.2cm 3.5cm 2.0cm 3.5cm},clip]{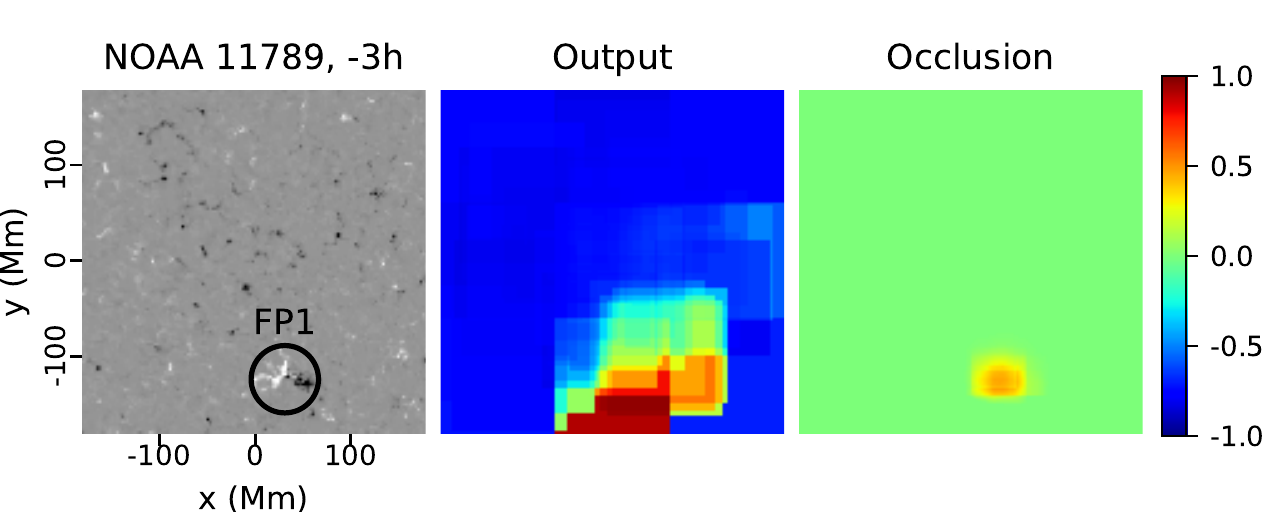}
\label{fig:visCR1}
}
\subfloat
{
\includegraphics[width=0.50\textwidth,trim={1.2cm 3.5cm 2.0cm 3.5cm},clip]{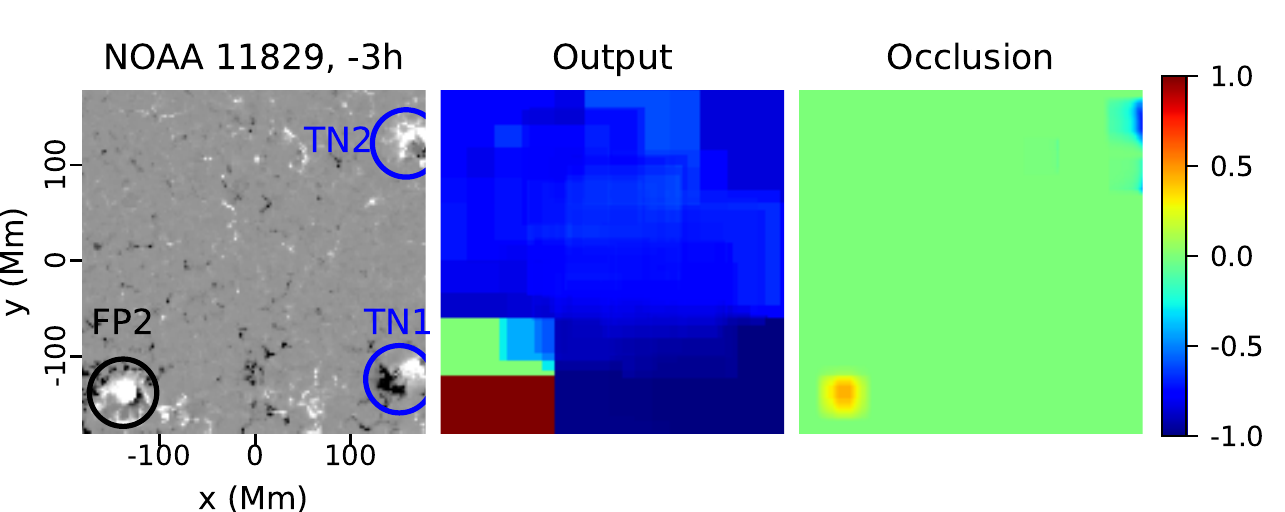}
\label{fig:visCR2}
}\\
\subfloat
{
\includegraphics[width=0.50\textwidth,trim={1.2cm 3.5cm 2.0cm 3.5cm},clip]{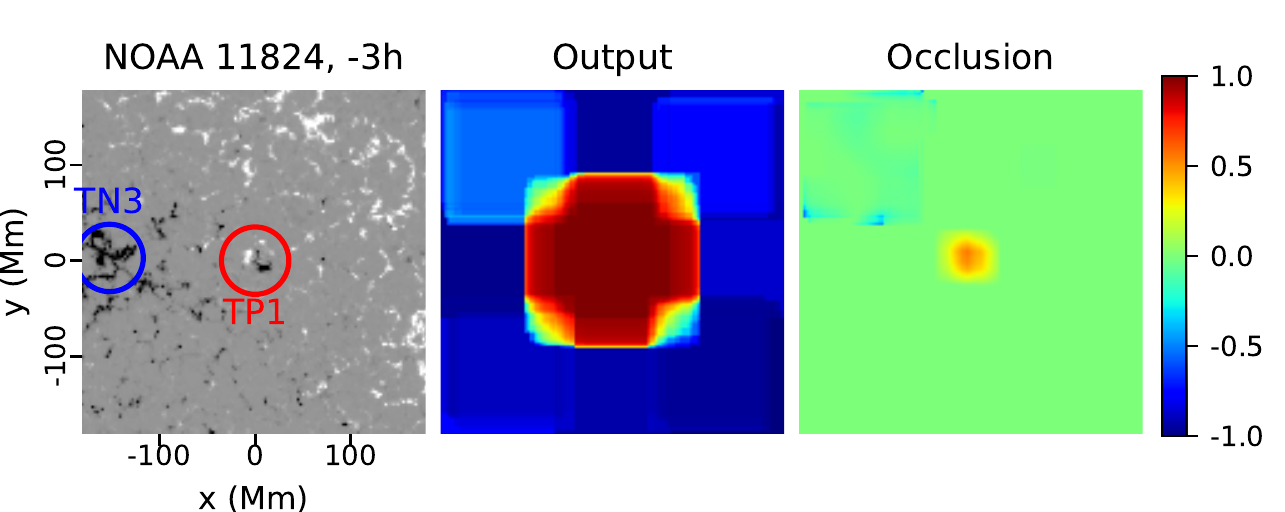}
\label{fig:visEAR1}
}
\subfloat
{
\includegraphics[width=0.50\textwidth,trim={1.2cm 3.5cm 2.0cm 3.5cm},clip]{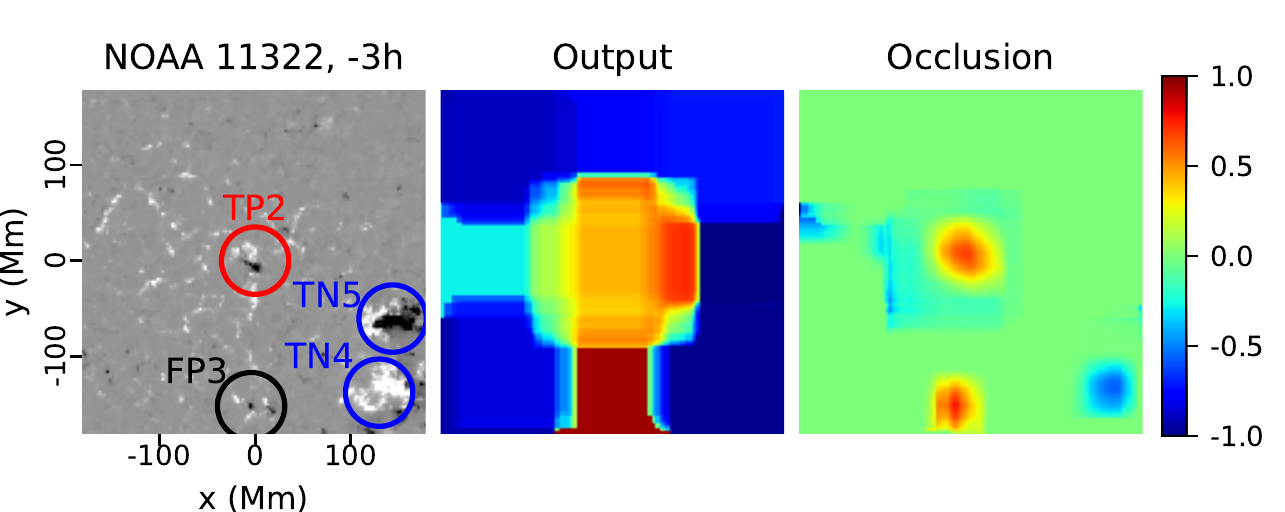}
\label{fig:visEAR2}
}
\caption{Visual depictions from the trained CNN using occlusion maps \citep{Zeiler2014}, obtained by systematically masking patches in the input and noting the change in the CNN output. The positive (negative) pixels in the occlusion map indicate regions in the input which are correlated (anti-correlated) with emergence as seen by the CNN. The CNN output and occlusion maps, calculated using Eqs.~\ref{eq:recombinedOutput} and \ref{eq:recombinedOcclusion} respectively, are shown for two control regions (CRs, {\it top}) and two emerging active regions (EARs, {\it bottom})) taken from 3.2~h before emergence. The magnetic field in EARs and CRs are saturated at 150 G (white) and -150 G (black). The output is saturated at 0 (blue) and 1 (red). The occlusion map is calibrated at -1 (blue) and 1 (red). Note that the emergence takes place within the central $10\degree \times 10\degree$ region of the EARs. Magnetic regions of interest, classified as true positives (TP, {\it red}), true negatives (TN, {\it blue}) and false positives (FP, {\it black}) are encircled.} 
\label{fig:visOccs}
\end{figure*}

\subsection{CNN output and the length-scale of magnetic regions.}
The aforementioned analysis shows that the outputs of the convolutional filters in the final layer strongly depend on the total unsigned line-of-sight flux $\Flux$ of PE and NE regions. $\Flux$ is also an important factor deciding the final CNN output, although there are other factors important for the classification (see Appendices~\ref{app:Bipolarity} and \ref{app:MTOT}). The value of $\Flux$ depends on the size of magnetic regions as well as the magnetic field intensity. We use synthetic magnetograms to explicitly test the dependence of the CNN output on both these factors.

We use circular synthetic bipolar magnetic regions with uniform magnetic field intensity, as shown in the left panel of Figure~\ref{fig:synBipole}, to probe the CNN output. The right panel of Figure~\ref{fig:synBipole} shows the pruned CNN output for synthetic bipoles of different magnetic field intensity and length-scales (radius). We see that the CNN output for a synthetic bipole of a given size generally increases with increasing value of the magnetic field intensity. The CNN output saturates for magnetic field value of $\sim 1000~\textrm{G}$ and falls sharply to $0$ beyond the length-scale of $\sim 30~\textrm{Mm}$. For the 100~G bipole, the CNN output peaks ($y \sim 0.9$) at the length-scale of 15~Mm and falls sharply to $0$ beyond the length-scale of $\sim 30~\textrm{Mm}$. For the 10~G bipole, the CNN output is consistently low ($y \sim 0.2$). Thus, the CNN associates magnetic bipoles of small length-scales and intense fields, as per Figure~\ref{fig:EAR-CR_Averaged}, with emergence.

Similar to Figure~\ref{fig:filterActVis}, Figure~\ref{fig:synFilterVis} shows outputs of the filters of the pruned CNN for a synthetic bipole. As discussed earlier, the filter outputs spatially coincide with the edges of the magnetic regions. Figure~\ref{fig:synFilters} shows contributions (maximum positive value times the filter weight, Eq.~\ref{eq:filterContri}) of the filters of the pruned CNN  as a function of the synthetic bipole length-scale. The contribution of filters 1 and 2, which are correlated with emergence, increases with increasing bipole length-scale upto $\sim 30~{\rm Mm}$ and saturates. In contrast, the contribution of filters 3 and 4, which are anti-correlated with emergence, continuously decreases with increasing bipole length-scale. No single filter is particularly sensitive to a specific length-scale of magnetic regions. The characteristic length-scale beyond which the CNN output decreases to $0$ is therefore a cumulative result of the contributions from all filters.

Figure~\ref{fig:synDepth} shows the dependence of the characteristic length-scale for the CNN output on the depth of the CNN (without pruning filters in the final layer). A correlation between the length-scale of the magnetic regions and the CNN output forms for a CNN with more than two layers which results in increased classification TSS (Table~\ref{tab:TSSwithDepth}). For CNNs with up to 5 layers, the length-scale beyond which the CNN output drops to $0$ increases with increasing number of CNN layers. This is result of downsampling of the input via max-pooling operation as the CNN captures patterns at increasing length-scales with increasing depth of the network. The characteristic length-scale continues to increase for CNN with layer 6 ($\sim 40~\rm{Mm}$).
after which it falls abruptly ($\sim 25~\rm{Mm}$). This is possibly a result of severe reduction of the input magnetograms (size $2 \times 2$ and $1 \times 1$ for layers 7 and 8 respectively) via max-pooling. Thus, the characteristic length-scale depends on downsampling of the input as a result of increasing number of CNN layers as well as the size of max-pooling kernel.

\subsection{CNN Visualizations.}
The analysis has thus far focused on deriving quantitative information about surface magnetic-field patterns that the CNN uses to discriminate between PE and NE magnetogram samples. The approach is specific to the problem at hand and has provided sufficient information to understand the operation of the trained CNN. The machine learning literature also provides us with tools for qualitative interpretation or visual explanations of deep neural networks \citep{Simonyan2013,Zeiler2014,Selvaraju2017}. Such tools, like saliency maps \citep{Simonyan2013} and gradient-based class-activation maps \citep{Selvaraju2017}, make use of the gradient information backpropagated \citep{hastie01statisticallearning} from the output layer to the hidden and input layers.

Here, we generate visualizations of the trained CNN using a simpler technique known as occlusion maps \citep{Zeiler2014}. Occlusion maps are obtained by measuring changes in the output when different patches in the input are masked. We use $25\times25$ pixel mask for input PE and NE magnetogram segments to generate occlusion maps. The mask size is chosen to be large enough to cause a significant difference in the CNN output after occlusion. For a given PE or NE sample, the occlusion map is of the same size as the magnetogram and initialized with the uniform value of the predicted class label $Y_{\rm pred} = 1\ {\rm or}\ 0$ for the sample. A new predicted class label $Y_{\rm mask}$ is obtained after masking or occluding a $25 \times 25$ pixel patch in the input image. The values at the corresponding pixels in the occlusion map are updated to $Y_{\rm pred} - Y_{\rm mask}$. Thus, the positive (negative) values in the occlusion map represent regions in the input image that are correlated (anti-correlated) with emergence as seen by the CNN. 
\begin{table}[t]
\centering
\begin{tabular}{cc|cc}
\toprule
&
$\Flux~(10^{20}~{\rm Mx})$ &&
$\Flux~(10^{20}~{\rm Mx})$\\
 \midrule
 TP1 & 10.8 & TN1 & 5.2\\
 TP2 & 15.2 & TN2 & 23.3\\
 FP1 & 7.2 & TN3 & 6.5\\
 FP2 & 55.2 & TN4 & 6.5\\
 FP3 & 8.6 & TN5 & 6.7\\
 \bottomrule
\end{tabular}
\caption{The total unsigned line-of-sight magnetic flux ($\Flux$) of $85 \times 85$-pixel segments including various regions of interests labeled in Figure \ref{fig:visOccs}.}
\label{tab:OccMTOT}
\end{table}

For comprehensive visualisation of occlusion maps corresponding to all PE and/or NE segments from an EAR or CR, we recombine sub-sampled magnetograms, corresponding CNN outputs and occlusion maps (see Appendix~\ref{app:reComb}). Figure~\ref{fig:visOccs} shows the CNN outputs and occlusion maps for two CR and two EAR samples. These samples are taken from 3.2~h before emergence. The $\Flux$ values of $85 \times 85$-pixel segments including various regions of interest are quoted in Table~\ref{tab:OccMTOT}. Note that, overall, $\Flux$ values of TN regions are low, with exception of TN2 for which $\Flux$ value is high. Also, overall, $\Flux$ values of TP and FP regions are intermediate, with exception of FP2 for which $\Flux$ value is very high. This is largely consistent with our earlier analyses. Regions TP1, TP2, FP1, FP2 and FP3 are highlighted positively in the occlusion map. Regions FP1 and FP3 show small-scale bipolar field patterns, which are an early sign of emergence, similar to regions TP1 and TP2. Regions TN2, TN3 and TN4 are highlighted negatively in the occlusion map, whereas, regions TN1 and TN5 are neutral in the occlusion map. These regions include bipoles, which may also be a factor for the CNN classification (Appendix~\ref{app:Bipolarity}). The CNN output $y\sim1.0$ for FP2 suggests that there may be other factors besides $\Flux$ and bipolarity for the CNN classification (see Appendix~\ref{app:MTOT}). These factors, however, are not easily interpreted.

\section{Summary} \label{sec:Discussion}
We perform a statistical analysis of emerging ARs in the \emph{SDO/HEAR} dataset \citep{Schunker2016} using convolutional neural networks (CNN). The trained CNN described here successfully classifies PE and NE magnetograms with a TSS of $\sim 85\%$, 3.2~h before emergence, which is significantly better than the TSS obtained in the LBB survey \citep{Leka_2012,Birch_2012,Barnes_2014}. The TSS score decreases to $\sim 40\%$, 24.5~h before emergence, similar to the LBB survey.

The discriminant analysis in the LBB survey \citep{Barnes_2014} showed that the pre-emergence surface magnetic field is a leading factor in distinguishing between EARs and CRs (Figure~\ref{fig:EAR-CR_Averaged}). We therefore perform a discriminant analysis of PE and NE magnetograms using the total unsigned line-of-sight magnetic flux $\Flux = \sum{\vert B_{\rm LOS} \vert~dA}$ to establish a baseline for comparing the CNN performance. We find that the maximum TSS achieved using $\Flux$ for classification of the PE and NE magnetograms is $\sim 65\%$, 3.2~h before emergence. Thus, the trained CNN performs better than the baseline model of discriminant analysis using only $\Flux$.

Through visual inspection of the PE and NE magnetograms, binned by the corresponding CNN output, we identify $\Flux$ as an important attribute of magnetograms for the CNN classification. We find that the CNN produces maximum output $y\sim1$ for magnetograms with $\Flux$ in an intermediate range. Magnetograms with very low as well as very high values of $\Flux$ yield low CNN output $y\sim0$. Further, statistical analysis of $\Flux$ of PE and NE magnetograms, also binned by CNN output, reveals that the average value of $\Flux$ for magnetograms which are predicted as emerging by the CNN, lies between $9\times10^{20}~\textrm{Mx}-11\times10^{20}~\textrm{Mx}$. These findings are validated by visual insights into the operation of the CNN obtained using occlusion maps \citep{Zeiler2014}. The TSS for classification decreases for PE and NE magnetograms from earlier pre-emergence times as the difference between distributions of $\Flux$ for the PE and NE samples diminishes (Figure~\ref{fig:EAR-CR_Averaged}). Explicit analysis of CNN output as a function of $\Flux$ shows that there are other factors besides $\Flux$ that contribute to the CNN performance (Appendix~\ref{app:MTOT}). These factors may be associated with length-scales, field intensities and geometry of the pre-emergence small-scale bipole shown in Figure~\ref{fig:EAR-CR_Averaged}.

The fully convolutional CNN (Figure~\ref{fig:CNN}) for this work was specifically chosen to facilitate subsequent assessment of the contribution of the convolutional filters in the final layer. Still, all filters in the final layer $N=256$ are difficult to analyse in detail to understand the specific patterns learned. We therefore develop a network-pruning algorithm to systematically remove unimportant filters from the final layer during training and retain only the most important (four) filters (Appendix~\ref{app:pruning}). From visual inspection as well as statistical analysis of filter outputs, we find that the CNN incorporates convolutional filters that are sensitive to total unsigned line-of-sight magnetic flux of PE and NE magnetograms. Some of these filters yield positive output in response to increasing magnetic flux, whereas others yield a negative output. The CNN output is the sum total of outputs from all filters. A multi-variable discriminant analysis of $\Flux$ and a measure of dipole moments of magnetograms yields the classification TSS of $75.6 \pm 3.3\%$ (3.2~h before emergence) which is comparable to the CNN (Appendix~\ref{app:Bipolarity}).

Using synthetic bipolar magnetograms of circular shape, we demonstrate that there exists a characteristic length-scale of magnetic regions beyond which the CNN output drops to $0$ irrespective of the magnetic field intensity. The characteristic length-scale is also a cumulative result of contributions from all filters and is an artifact of the number of layers in the CNN and max-pooling operation. Explicit dependence on these factors remains to be explored. 

The CNN output peaks for small-scale ($\sim 30 ~{\rm Mm}$) and intense ($\sim 100~{\rm G}$) synthetic bipoles. Our analysis does not conclusively show that bipolarity of magnetograms is an important factor for the CNN classification. Interestingly, a CNN trained on PE and NE segments without any polarity information also yields a comparable value of the TSS (Appendix~\ref{app:ABSData}). Altogether this suggests that the CNN identifies small-scale and (unipolar or bipolar) intense fields as a characteristic pre-emergence pattern. A time-dependent analysis is necessary to shed light on formation of such small-scale fields by either rising intact from deep within the convection zone \citep{Fan2009} or by forming over a less localized area near surface \citep{Brandenburg_2005}. Further, \citet{Birch2019} show that the emerging flux interacts with the supergranulation pattern and therefore emergence locations are correlated with supergranulation. The network is highlighted in PE and NE magnetograms since magnetic flux tends to collect in supergranular lanes. Thus, the CNN performance here may also be dependent on factors associated with the supergranulation. These remain to be explored.

It is to be noted that the \emph{SDO/HEAR} dataset considered here contains limited ($\sim 200$) independent samples of EARs and CRs. The quantity of data required for training the CNN to gain deeper and concrete insights about pre-emergence patterns may not be known in advance. However, the limited dataset hinders the CNN from learning the convolutional filters associated with more and general complex patterns than just the total unsigned pre-emergence magnetic field, e.g., small-scale pre-emergence bipoles. The SDO/HMI instrument provides terabytes of high resolution, high cadence magnetic field data every day. With these data, it may be possible to construct a large enough dataset, with more general selection criteria compared to the \emph{SDO/HEAR} dataset, to study emergence. Such a dataset is also likely to conform to real-time emergence scenarios and therefore may also be analysed with a focus on forecasting emergence. Alternatively, it may be possible to train the CNN convolutional filters to exactly detect the pre-emergence bipolar magnetic-field and correlated patterns using additional constraints on convolutional filters \citep{Bayar2018} and/or to learn the emergence location along with a probability \citep{zhao2018object}. A machine successful for such a task will also be useful for the analysis of pre-emergence temporal evolution. These analyses are deferred to future work. With Petabyte-scale astronomical datasets expected in the upcoming decade, machine learning is expected to become increasingly relevant in analysing data \citep{baron2019machine}. Insights obtained from this work may therefore be useful for training and interpreting deep neural networks in future solar and astrophysical data-analysis applications.

\acknowledgments
D.B.D., S.M.H., A.C.B. and H.S. designed the research. D.B.D. performed the data analysis. All authors contributed to the interpretation of the results. D.B.D. wrote the manuscript with contributions from all authors.

S.M.H acknowledges funding support from the Max-Planck partner group program. The HMI data used here are courtesy of NASA/SDO and the HMI science team. We acknowledge partial support from the European Research Council Synergy Grant WHOLE SUN $\#$810218. Observations are courtesy of NASA/SDO and the HMI science teams. We thank the anonymous reviewer for their comments and suggestions which helped improve the analysis and clarity of the paper.


\appendix

\section{Neural Networks. \label{app:NNs}} 
\begin{figure}[b]
\centering
\includegraphics[width=0.60\textwidth,trim={15cm 7cm 12cm 7cm},clip]{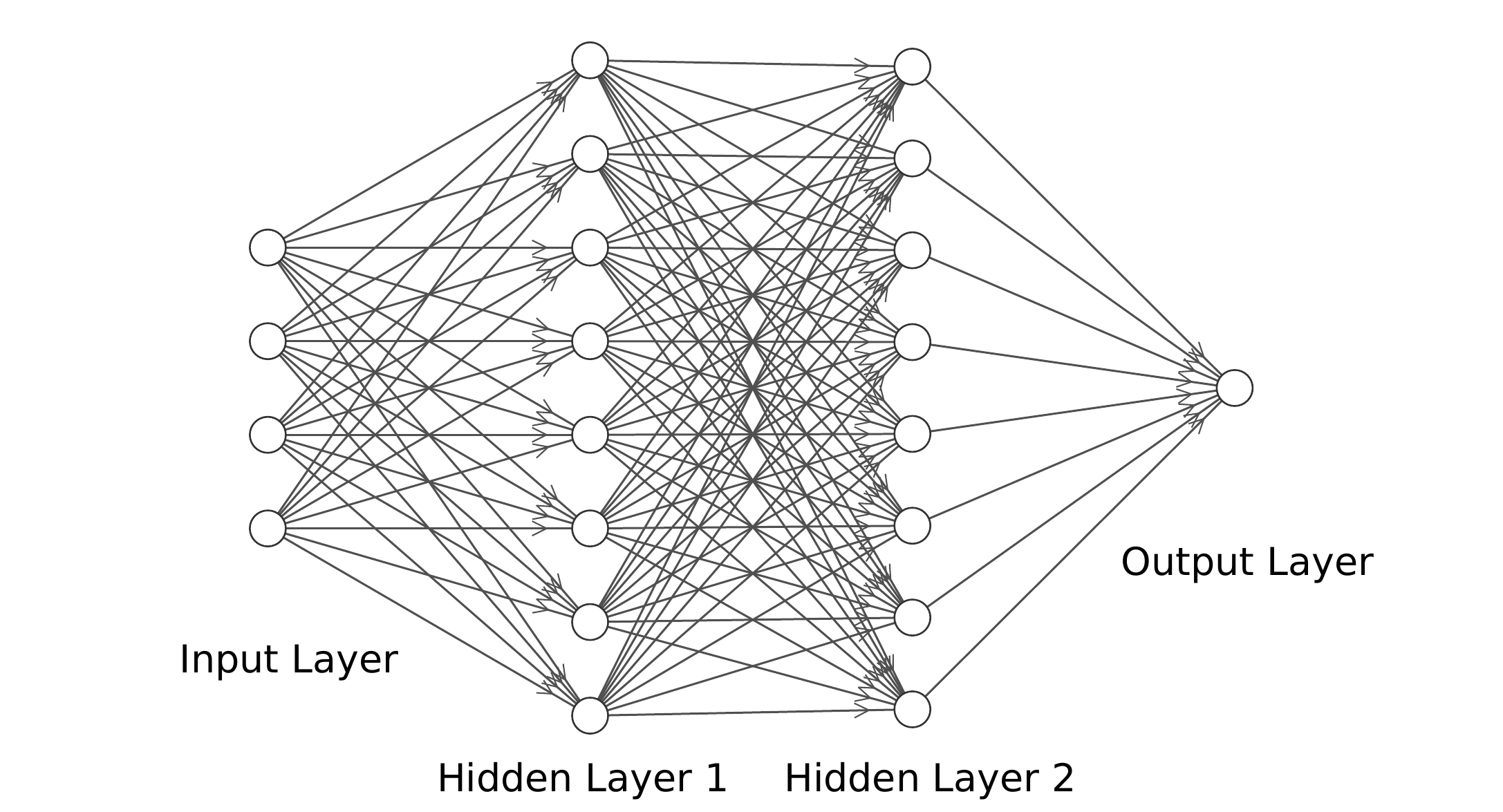}
\caption{A two-layer deep fully connected (FC) neural network with four inputs and one output,}
\label{fig:NN_Illustration}
\end{figure}
A deep neural network consists of many layers of neurons as shown in Figure~\ref{fig:NN_Illustration}. The $i^{\rm th}$ neuron in the $n^{\rm th}$ layer performs the following generic operation
\begin{equation} \label{eq:neuronAct} 
    x_i^{n} = f(\sum_{j=1}^{N} w_{ij}^{n} x_j^{n-1} + b_i^{n}).
\end{equation}
Here, $\textbf{x}^{n-1}$ are the outputs from the neurons in the $(n-1)^{\rm th}$ layer (the $0^{\rm th}$ layer is the input layer), $w_{ij}^{n}$ is the weight for the $j^{\rm th}$  input from the $(n-1)^{\rm th}$ layer, $b_i^{n}$ is termed the bias and $f$ is the activation function for the neuron. The activation applied on the final layer produces the output $y^{\rm pred}$. The error in the predicted output is determined by a loss function $L(y,y^{\rm pred})$. During training, weights and biases of the neurons in the network are determined via stochastic gradient descent to minimize the loss $L(y,y^{\rm pred})$. Here, we use binary cross-entropy (CE) as the loss function, which is a popular choice for classification problems \citep{hastie01statisticallearning}. It is defined as
\begin{equation} \label{eq:LossFunction} 
    L_{\rm CE}(y,y^{\rm pred}) = -\frac{1}{N} \sum_{i=1}^{N} \left[ y_{i} log \left(y^{\rm pred}_i\right) + \left(1-y_i\right) log \left(1-y^{\rm pred}_i\right) \right], 
\end{equation}
where $N$ is the batch-size, i.e., the number of examples considered in the stochastic gradient descent. The training process involves many cycles of iterations (epochs) over the entire training set, and loss function $L_{CE}$ converges gradually during training.

\section{sub-population of PE and NE samples. \label{app:subpop}}
\begin{figure}[t]
\centering
\includegraphics[width=0.5\textwidth,trim={0.2cm 0.2cm 0.2cm 0.2cm},clip]{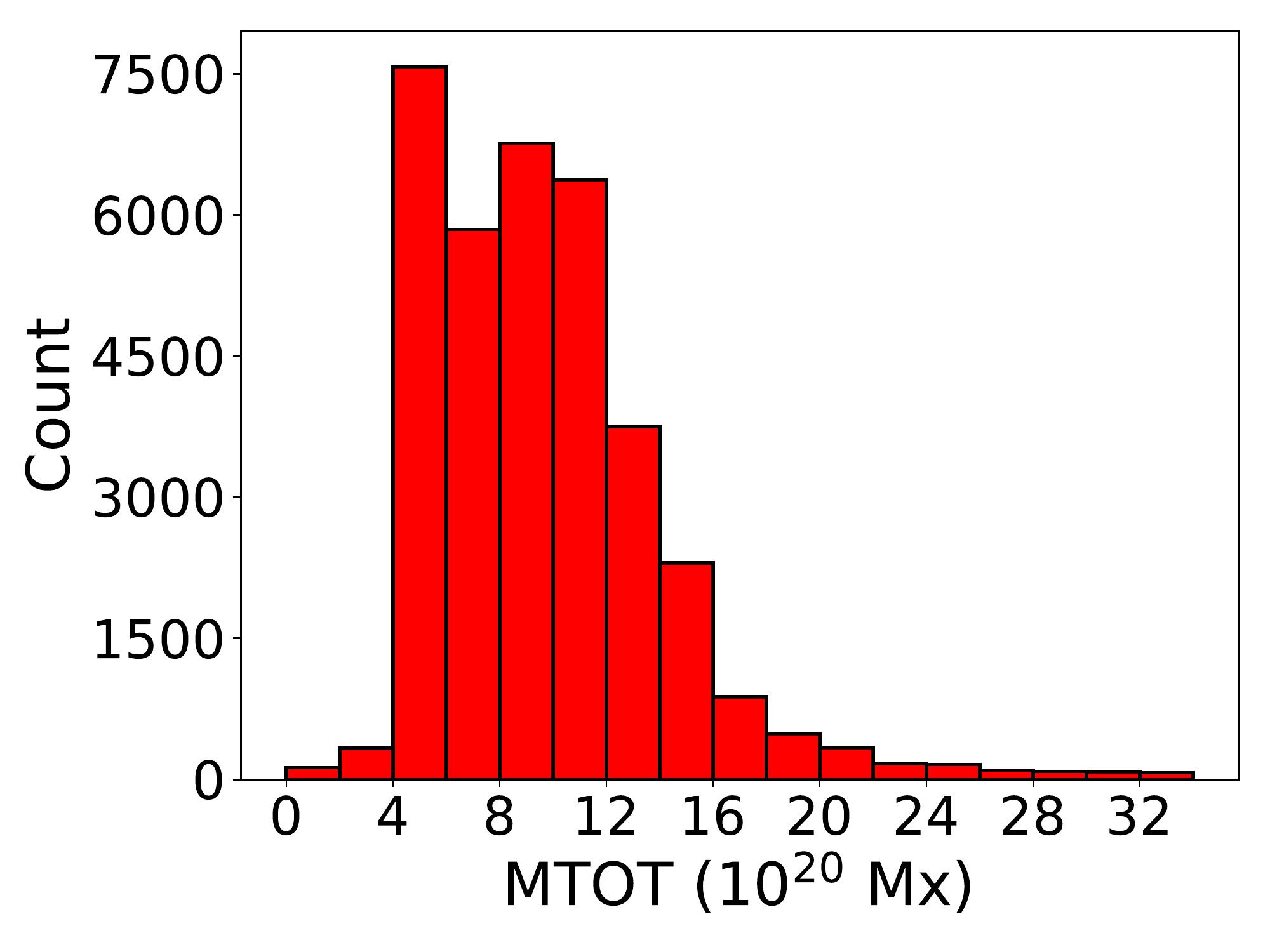}
\caption{Distribution of pre-emergence (PE) and non-emergence (NE) samples taken at 3.2~h before emergence according to the total unsigned line-of-sight magnetic flux ($\Flux$). We select samples with $4.0 \times 10^{20}~{\rm Mx} \leq $\Flux$ \leq 16.0 \times 10^{20}~{\rm Mx}$ for the statistical analysis.
}
\label{fig:subpopulation}
\end{figure}

The attribute of the PE and NE magnetograms that stands out for the CNN classification is the total unsigned line-of-sight magnetic flux $\Flux$ (Section~\ref{sec:Results}). Therefore, we analyse $\Flux$ of PE and NE samples from each category of the CNN output bins. The distribution of PE and NE samples as per $\Flux$ is shown in Figure~\ref{fig:subpopulation}. For the statistical analysis binned as per $\Flux$ (Figure~\ref{fig:filterActVis}), it is desirable that each bin contains sufficient number of samples for meaningful statistics. We therefore select samples with $\Flux$ value between $4 \times 10^{20}~{\rm Mx}$ and $16 \times 10^{20}~{\rm Mx}$ (both inclusive). For each bin within this $\Flux$ range, number of samples available are large $>1500$.

\section{Network pruning \label{app:pruning}}
Neural network pruning was primarily proposed to reduce the computation and memory required for tasks such as vision and speech recognition or natural language processing on embedded mobile applications \citep{Han2015}. The pruned network, significantly reduced in size compared to the original, is also easier to interpret \citep{Frankle2019}. A straightforward approach to network pruning is to identify important neuron connections during the training step and remove all connections with weights below a threshold value \citep{Han2015}. This process may be repeated in an iterative manner during training to obtain a compressed trained network. In the functionality based approach, one can remove the neurons from the network that are functionally similar to the other neurons \citep{Qin2018}. 

Here, we perform pruning of the final convolutional layer of the CNN, in an iterative manner, to reduce the number of filters from 256 to 4. Our pruning strategy is based on identifying the convolutional filters that maximally contribute towards the network output. In the CNN (Figure~\ref{fig:CNN}), the final convolutional layer is connected to the output layer through a $6\times6$ max-pooling layer. The max-pooling layer picks out the pixel with the maximum value from the $6\times6$ output of a convolutional filter. Thus for a given input, the contribution of each filter in the final layer is given by
\begin{equation} \label{eq:filterContri}
    F_{i} = max(O_{i})w_i, 
\end{equation}
where $O_i$ is the $6\times6$ output of the $i-th$ convolutional filter in the final convolutional layer and $w_i$ is the corresponding weight. The output of the CNN is obtained by applying {\it sigmoid function} \citep{hastie01statisticallearning} to the sum of the output of each convolutional filter 
\begin{equation} \label{eq:CNNoutput}
    y=sigmoid(\sum_{i=1}^{N}{F_i} + b),
\end{equation}
where $b$ is the bias of the output neuron and $N$ is the number of convolutional filters in the final layer. The output $y$, a value between 0 and 1, is a measure of the probability of the input magnetogram showing emergence in a certain time $t$. The contribution of CNN filters $F_{i}$, which correspond to the surface magnetic-field patterns correlated (anti-correlated) with emergence, increases (decreases) the CNN output $y$.

We develop a pruning algorithm to retain top four filters while maintaining prediction accuracy --- two each corresponding to patterns correlated and anti-correlated with emergence. We initiate the training with $N=256$ filters in the final convolutional layer and perform the pruning operation during training, repeatedly applied every few epochs. At each pruning step, half the number of filters associated with patterns minimally correlated and anti-correlated with emergence are removed. 

We identify the filters to be pruned in the following manner. At each pruning step during training, we calculate the total contribution $S_{i}^{\rm TP}$ of each convolution filter $i$ for all samples $n^{\rm TP}$ predicted as true positives (TPs) from the training data
\begin{equation} \label{eq:TPContri}
    S_{i}^{\rm TP} = \sum_{j=1}^{n^{\rm TP}} {F_i}.
\end{equation}
Similarly, we calculate the total contribution $S_{i}^{\rm TN}$ for all samples  predicted as true negatives (TNs) from the training data.
Finally, we calculate net positive contribution $S_{i}^{\rm net}$ of each convolutional filter $i$ for all accurately classified samples from the training data
\begin{equation} \label{eq:netContri}
    S_{i}^{\rm net} = S_{i}^{\rm TP} - S_{i}^{\rm TN}.
\end{equation}
\begin{figure}[b]
\centering
\includegraphics[width=0.45\textwidth,trim={0cm 0cm 0cm 0cm},clip]{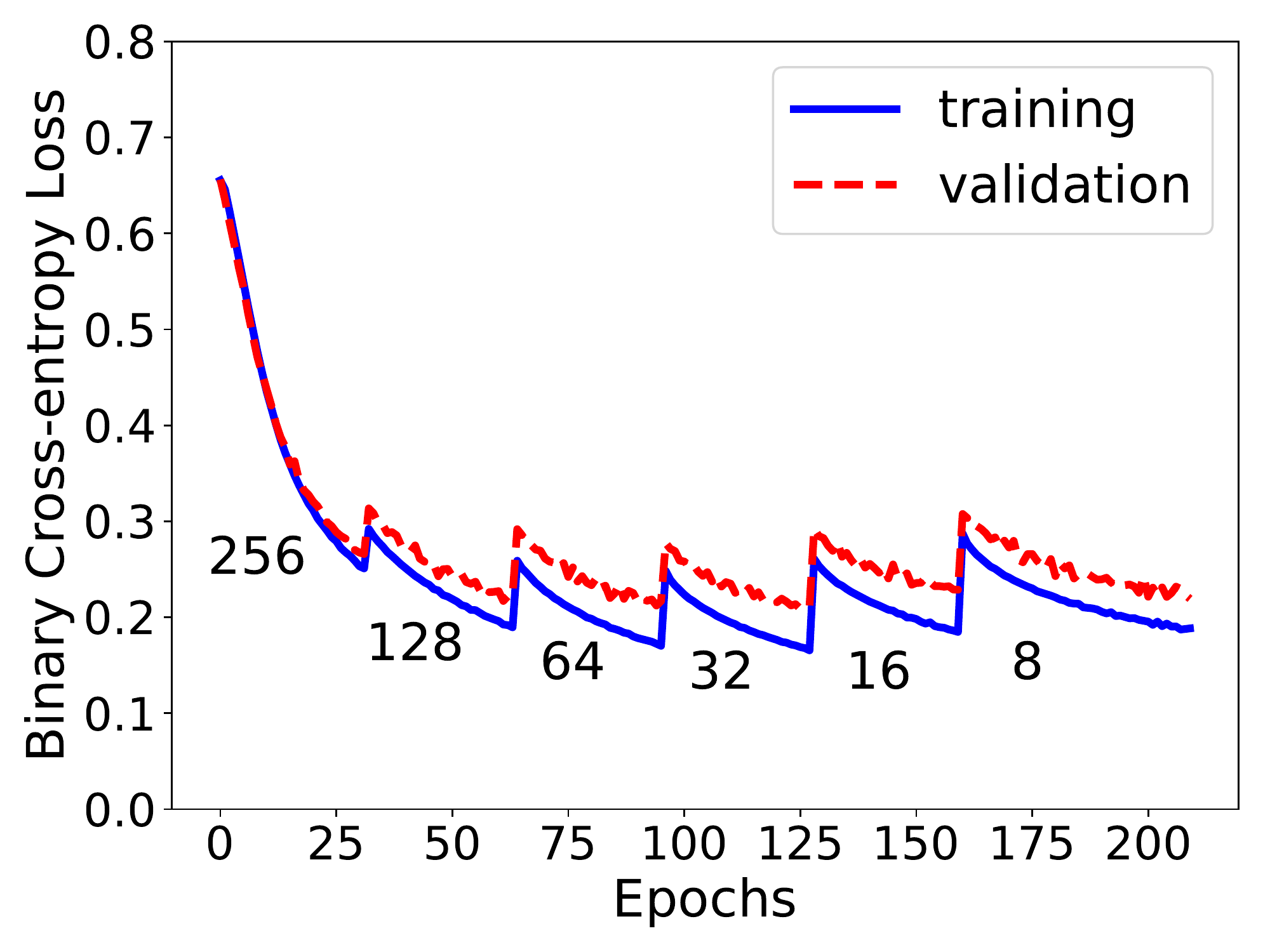}
\caption{Convergence of training and validation loss (Eq.~\ref{eq:LossFunction}) during training of the CNN using network pruning. At a pruning stage, half the number of the CNN filters from the final convolutional layer, which are least correlated with the surface field patterns associated with emergence, are removed. The pruning shows up as discontinuities in the training and validation losses. The number of CNN filters in the final convolutional layer at each training stage are indicated below the curve for that corresponding stage. Pruning is carried out after every 32 epochs, except at the last stage, when it is allowed to run for 50 epochs. The final pruning step is performed at the end of the training, yielding the trained CNN with 4 convolutional filters in the final layer. The CNN is not retrained subsequent to the final pruning. We use slightly higher learning rate, $lr = 5 \times 10^{-7}$.}
\label{fig:prunedCNNTraining}
\end{figure}
For a convolutional filter associated with patterns maximally correlated with emergence $S_{i}^{\rm net} >> 0$. Likewise, for a convolutional filter associated with patterns maximally anti-correlated with emergence $S_{i}^{\rm net} << 0$. Thus, at each pruning step, we retain $N/4$ filters with maximum $S_{i}^{\rm net}$ value and $N/4$ filters with minimum $S_{i}^{\rm net}$ value. The remaining $N/2$ filters, which are relatively neutral to the patterns associated with emergence, are removed. The network is trained further for a pre-defined number of epochs, after which it is pruned again to remove half the number of neutral filters. This is repeated until only four filters remain in the final layer --- the dominant two corresponding each to patterns correlated and anti-correlated with emergence. The step by step algorithm is as follows.

\begin{enumerate}[noitemsep]
    \item Initialize the CNN (Figure~\ref{fig:CNN}) with the final convolutional layer filters $N=256$.
    \item Initialize a function for the number of training epochs $E(N)$ 
    as a function of the number of the final convolutional layer filters.
    \item Train the network for epochs $E(N)$.
    \item Using the trained network, obtain TP samples $n^{\rm TP}$ and TN samples $n^{\rm TN}$ of all PE and NE magnetograms in the training data. 
    \item Calculate the net contribution of each of the $i$ filters $S_{i}^{\rm net}$ (Eq.~\ref{eq:netContri}) towards the CNN output for all accurately predicted PE and NE samples (TPs and TNs) in the training data.
    \item Identify $N/4$ filters with maximum (positive) $S_{i}^{\rm net}$ and $N/4$ filters with minimum (negative) $S_{i}^{\rm net}$. Remove the remaining $N/2$ filters.
    \item  {\it If} $N=4$, i.e. after pruning, stop and output the trained CNN model. {\it Else} go to 3.
\end{enumerate}

Using the aforementioned algorithm, we train the CNN and iteratively prune the final convolution filters from $N=256$ to $N=4$. Training and validation losses as training progresses are shown in Figure~\ref{fig:prunedCNNTraining}. We see that the losses converge in an identical manner, with discontinuities encountered at each pruning stage. The number of filters in the final convolutional layer is shown below the curve corresponding to each pruning stage. At each pruning step, half the convolutional filters are removed according to the pruning algorithm. At the end of the final epoch, pruning is carried out for the last time, reducing the convolution filters from $N=8$ to $N=4$. The network is not retrained subsequently.

The CNN, with the final convolution layer pruned to $N=4$ filters, yields ${\rm TSS}=82.50\% \pm 4.52\%$ for classification of PE and NE samples (-3.2~h pre-emergence) which is comparable to the original trained CNN with $N=256$ filters (Table \ref{tab:TSSwithTime}). The CNN obtained from pruning a random assortment of $N/2$ filters in the final convolution layer at each step yields ${\rm TSS}=34.71\% \pm 28.86\%$. This validates the algorithm developed for identifying filters in the final convolutional layer correlated and anti-correlated with surface magnetic-field patterns associated with emergence. We analyse the top four filters that are retained after the pruning to interpret the performance of the CNN for classification of the PE and NE magnetograms (Section~\ref{sec:Results}). Note that, in this case, the CNN with $N=4$ filters in the final convolutional layer from the outset and trained for an identical number of epochs yields ${\rm TSS} = 78.38\% \pm 6.60\%$ which is also comparable to the original trained CNN. Since, in general, the number for optimum filters (neurons) in a CNN (deep neural network) is not known in advance, the functionality based pruning approach developed here is useful for future applications as well. Further pruning the CNN to $N=2$ filters in the final layer yields a mean TSS of 68.9\% with a significantly larger standard deviation of 17.1\%.

\section{Recombining sub-sampled PE and NE magnetograms. \label{app:reComb}}
The $10\degree \times 10\degree$ PE and NE magnetograms used for training the CNN are randomly sub-sampled from $30\degree \times 30\degree$ EARs and CRs. Hence, PE and/or NE magnetograms sub-sampled from an EAR or CR may contain overlapping regions (see Section~\ref{sec:Data}). These PE and/or NE segments sub-sampled from an EAR or CR may be recombined to recover the original EAR or CR in the following manner. Let the $(i,j)$ pixel from an EAR or CR be included in $C_{ij}$ number of PE and/or NE magnetogram samples from the EAR or CR. We obtain the net CNN output corresponding to each pixel of the EAR or CR as 

\begin{equation} \label{eq:recombinedOutput}
    y_{ij} = (1/C_{ij}) \sum_{k=1}^{C_{ij}} {y_{ij}^{k}},
\end{equation}
where ${y_{ij}}^{k}$ is the CNN output corresponding to $k-th$ PE and/or NE sample from the EAR or CR. Similarly, the occlusion map $A$ for each EAR or CR pixel is obtained as
\begin{equation} \label{eq:recombinedOcclusion}
    A_{ij} = (1/C_{ij}) \sum_{k=1}^{C_{ij}} {A_{ij}^{k}},
\end{equation}
where ${A_{ij}}^{k}$ is the occlusion map corresponding to the $k-th$ PE and/or NE sample from the EAR or CR. By recombining PE and/or NE magnetogram segments into the original EAR or CR, we comprehensively inspect outputs and occlusion maps of all PE and/or NE magnetograms that are part of the respective EAR or CR.

\section{Multi-variable discriminant analysis of the line-of-sight unsigned magnetic flux and dipole moment. \label{app:Bipolarity}}
\begin{figure}[t]
\centering
\includegraphics[width=0.45\textwidth,trim={0cm 0cm 0cm 0cm},clip]{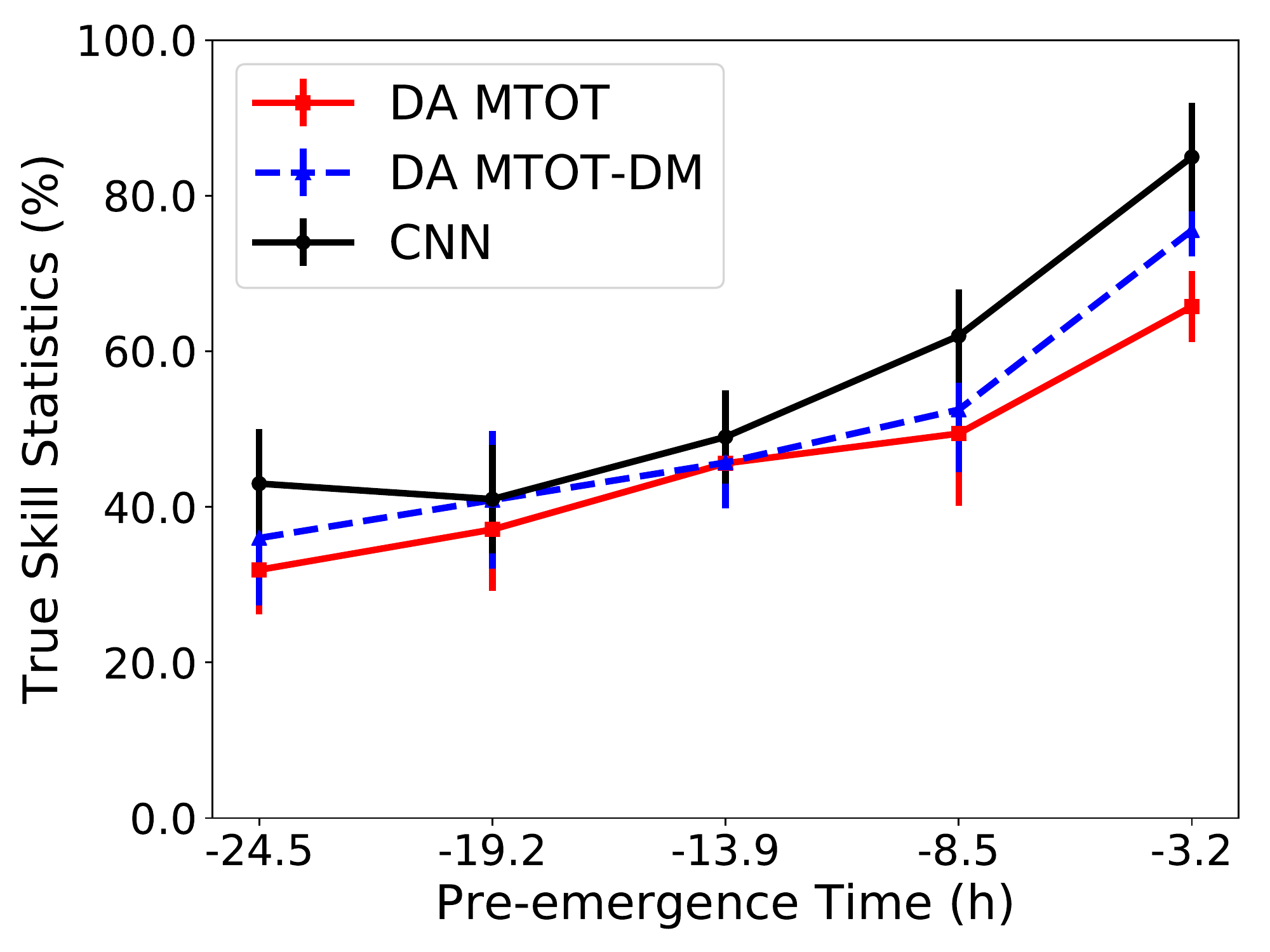}
\caption{Comparison of TSS obtained using non-parametric multi-variable discriminant analysis of the total unsigned line-of-sight flux ($\Flux$) and a measure of dipole moment (DM, \ref{eq:DM}) at different pre-emergence times. The classification TSS of multi-variable discriminant analysis is comparable with the classification TSS of the CNN and significantly better than the TSS obtained using discriminant analysis of only $\Flux$. The $1\sigma$ error bars are shown.}
\label{fig:BPDA}
\end{figure}
Through visual inspection and statistical analysis of accurately classified PE and NE magnetograms, we find that $\Flux$ is an important factor for the CNN classification. As noted in Figure~\ref{fig:EAR-CR_Example}, appearance of a small-scale bipolar field is an early sign of emergence. Therefore, bipolarity of the magnetograms may also be a factor important for the CNN classification.

We calculated a baseline TSS using a non-parametric discriminant analysis of only the unsigned line-of-sight magnetic flux ($\Flux$). We extend the baseline model to include a measure of `dipole moment' of the magnetograms \citep{Illarionov2015,wilson_1994} calculated as 
\begin{equation} \label{eq:DM}
    {\rm DM} = \left\vert{\sum_{i} B_{i} \left(\bf{r}_i - \bf{r}_0\right)}\right\vert,
\end{equation}
where $B_{i}$ is the magnetic field of the $i^{th}$ pixel located at $\bf{r}_i$, $\bf{r}_0$ is a reference pixel and the sum is over all pixels in the magnetogram. We estimate probability density of PE and NE magnetograms as a function of two variables --- $\Flux$ and DM --- using the Epanechnikov kernel \citep{Silverman86,Barnes_2014}. Similar to the non-parametric discriminant analysis of $\Flux$, the smoothing parameter for the density estimation is chosen to be optimum for a normal distribution. For each cross-validation set, probability density for PE and NE magnetograms is estimated using the training data. The trained density estimator is then used to obtained the probability score for PE and NE magnetograms in the validation data and a classification label is obtained for calculating the TSS. Cross-validation TSS for the multi-variable discriminant analysis for different pre-emergence times thus obtained is plotted in Figure~\ref{fig:BPDA}. We see that the multi-variable discriminant analysis using $\Flux$ and DM is comparable to the CNN classification results considering the error bars. 

\section{CNN classification of uniform positive polarity PE and NE magnetograms. \label{app:ABSData}}
Our analysis shows that the CNN classification depends on $\Flux$ and outperforms the discriminant analysis of only $\Flux$. We also show that the discriminant analysis of $\Flux$ and DM --- a measure of bipolarity --- is comparable to the CNN classification. This suggests that the bipolarity of magnetograms may also be an important factor for the CNN classification. To validate this, we train the CNN (Figure~\ref{fig:CNN}) with uniform positive polarity PE and NE magnetograms i.e. the absolute value of the line-of-sight field. We find that this CNN yields TSS comparable to the CNN trained on original PE and NE magnetograms (see Table~\ref{tab:ABSTSS}). Therefore, the polarity information in the data is not necessary for the CNN to yield the classification performance superior to the discriminant analysis of only $\Flux$.

\begin{table}[b]
\centering
\begin{tabular}{cc}
\toprule
time before  & 5-fold cross-validation \\
emergence (h) & TSS ($\%$)\\
 \midrule
-3.2 & 88.20 $\pm$ 4.78  \\
-8.5 & 60.07 $\pm$ 1.53 \\
 -13.9 & 50.66 $\pm$ 7.32 \\
 -19.2 & 43.89 $\pm$ 7.30 \\
-24.5 & 40.68 $\pm$ 2.03 \\
 \bottomrule
\end{tabular}
\caption{Mean 5-fold cross-validation {\it True Skill Statistics} (TSS) for classification of uniform-polarity pre-emergence (PE) and non-emergence (NE) magnetograms at different pre-emergence times obtained using the convolutional neural network (CNN, Figure~\ref{fig:CNN}) with 5 hidden convolutional layers. The $1\sigma$ error is quoted.}
\label{tab:ABSTSS}
\end{table}

\begin{figure}[t]
\centering
\includegraphics[width=\textwidth,trim={0.0cm 4.8cm 0.0cm 4.5cm},clip]{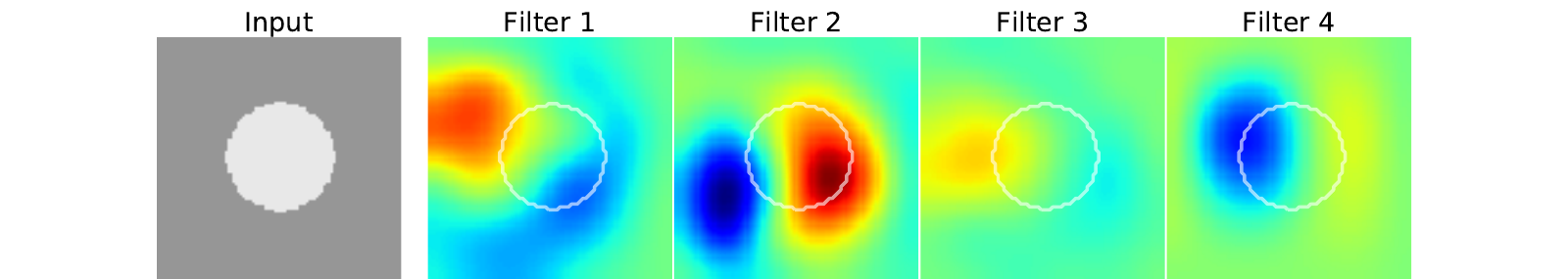}
\caption{Outputs of filters in the final convolutional layer of the CNN, with $N=4$ filters in the final layer, trained on uniform-polarity PE and NE segments. The input is a synthetic magnetogram with 100~G uniform-polarity field and radius 25~Mm.}
\label{fig:ABSsynFilterVis}
\end{figure}

To understand the operation of the CNN trained on uniform-polarity data, we train a CNN with $N=4$ filters from the outset in the final layer and visualize the filter outputs in the final layer corresponding to a synthetic magnetogram with uniform field similar to Figure~\ref{fig:synFilterVis}. The CNN with $N=4$ filters in the final layer yields TSS of $71.75 \pm 8.68 \%$. The final layer filter outputs are shown in Figure~\ref{fig:ABSsynFilterVis}. We find that the filter outputs conform to the edges of magnetic-flux regions and are both positive as well as negative. Since the synthetic magnetogram contains no information about the polarity, the positive and negative filter outputs do not necessarily correspond to the polarity of the magnetic-flux regions in Figure~\ref{fig:ABSsynFilterVis} and also in Figure~\ref{fig:synFilterVis}. The filter outputs are a result of complex operations performed by the CNN in the hidden layers and may not be easily interpreted.

\section{CNN output vs $\Flux$. \label{app:MTOT}}
\begin{figure}[t]
\centering
\includegraphics[width=0.40\textwidth,trim={0.0cm 0cm 0.0cm 0cm},clip]{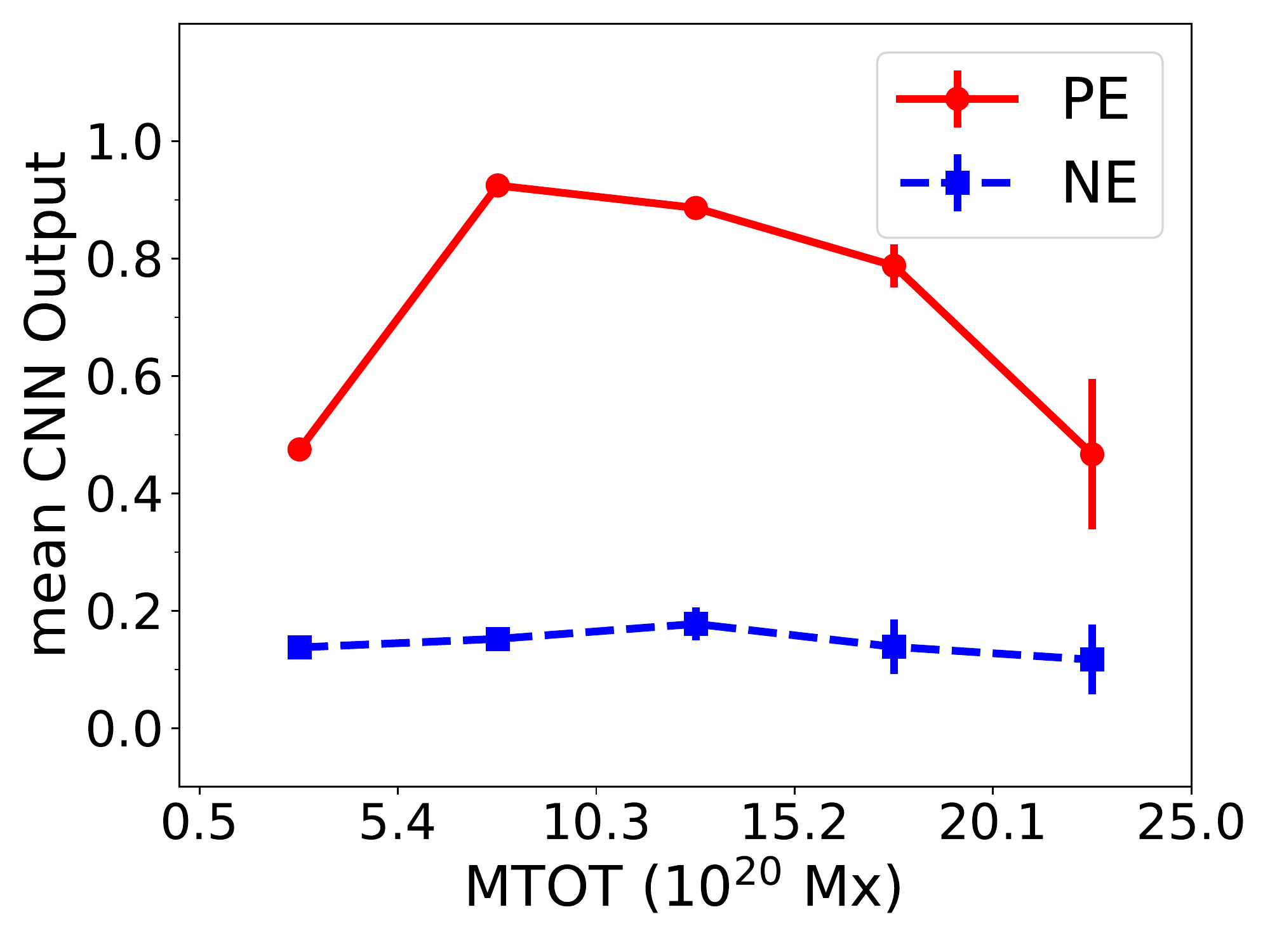}
\caption{The average CNN output $\langle y \rangle$ of PE and NE magnetogram samples binned by the unsigned line-of-sight magnetic flux $\Flux$. PE and NE samples considered are from $\Flux$ range of $0-25 \times 10^{20} {\rm Mx}$. ${\langle y \rangle}_{\rm NE}$ is consistently lower than 0.2. ${\langle y \rangle}_{\rm PE}$ is greater than 0.5 for an intermediate range of $\Flux$ between $5 \times 10^{20} -20 \times 10^{20} {\rm Mx}$. 5$\sigma$ error bars are shown.}
\label{fig:binnedOutput}
\end{figure}

In Figure~\ref{fig:SA}, we see that the average unsigned line-of-sight magnetic flux $\langle\Flux\rangle$ of PE and NE samples binned by the CNN output lies in an intermediate range of $9 \times 10^{20} - 11 \times 10^{20}~{\rm Mx}$ for samples with the CNN output $y \sim 1$. While this indicates a dependence of the CNN output on $\Flux$, the actual values of $\Flux$ of samples with $y \sim 1$ may not all lie in the intermediate range and may be higher as well as lower. To explicitly study the dependence of the CNN output on $\Flux$, we consider the average CNN output of PE and NE samples binned by $\Flux$. We consider samples from a slightly wider range of $\Flux$ of $0-25 \times 10^{20}~{\rm Mx}$ than considered earlier (Appendix~\ref{app:subpop}). The average CNN output $\langle y \rangle$ is plotted in Figure~\ref{fig:binnedOutput}. While ${\langle y \rangle}_{\rm NE}$ is consistently low $<0.2$ and mostly independent of $\Flux$, ${\langle y \rangle}_{\rm PE}$ is $>0.5$ for samples with $\Flux$ values between $5 \times 10^{20}-20 \times 10^{20}~{\rm Mx}$ and $<0.5$ for samples with lower or higher $\Flux$ values. Importantly, ${\langle y \rangle}_{PE}$ is consistently and significantly higher than ${\langle y \rangle}_{NE}$ irrespective of $\Flux$. This suggests that the CNN output depends on other important factors besides $\Flux$.   

\end{document}